\documentclass[onecolumn,tighten,times]{aastex62}
\usepackage{graphicx}
\usepackage{color}
\usepackage{mathrsfs}
\usepackage{color,ulem}                                                                                                                                        
\usepackage{float}                                                                                                                                             
\usepackage{subfigure}                                                                                                                                         
\usepackage{amsmath}
\usepackage{empheq}
\usepackage{bm}
\usepackage{mathrsfs}
\usepackage{ulem}

\def\mbh{M_{\bullet}}

\def\ergs{\rm \ erg~s^{-1}}
\def\ergscma{\rm \  erg~s^{-1}~cm^{-2}~\AA^{-1}}
\def\fblr{f_{_{\rm BLR}}}

\def\Heii{He {\sc ii}}
\def\kms{\rm \  km~s^{-1}}

\def\mathdotM{\dot{\mathscr{M}}}

\def\pp{\prime\prime}

\def\sunm{M_{\odot}}

\def\mcn{\multicolumn}

\def\feii{Fe {\sc ii}}

\def\oiii{[O~{\sc iii}]}

\newcommand{\sersic}{S\'{e}rsic}

\begin{document}


\title{REVERBERATION MAPPING OF NARROW-LINE SEYFERT 1 GALAXY I ZWICKY 1: BLACK HOLE MASS}

\author{Ying-Ke Huang}
\affiliation{Key Laboratory for Particle Astrophysics, Institute of High Energy Physics, Chinese Academy of Sciences, 19B Yuquan Road, Beijing 100049, China}
\affiliation{School of Astronomy and Space Science, University of Chinese Academy of Sciences, 19A Yuquan Road, Beijing 100049, China}

\author{Chen Hu}
\affiliation{Key Laboratory for Particle Astrophysics, Institute of High Energy Physics, Chinese Academy of Sciences, 19B Yuquan Road, Beijing 100049, China}

\author{Yu-Lin Zhao}
\affiliation{Kavli Institute for Astronomy and Astrophysics, Peking University, Beijing 100871, China}
\affiliation{Department of Astronomy, School of Physics, Peking University, Beijing 100871, China}

\author{Zhi-Xiang Zhang}
\affiliation{Key Laboratory for Particle Astrophysics, Institute of High Energy Physics, Chinese Academy of Sciences, 19B Yuquan Road, Beijing 100049, China}
\affiliation{School of Astronomy and Space Science, University of Chinese Academy of Sciences, 19A Yuquan Road, Beijing 100049, China}

\author{Kai-Xing Lu}
\affiliation{Yunnan Observatories, Chinese Academy of Sciences, Kunming 650011, China}

\author{Kai Wang}
\affiliation{Key Laboratory for Particle Astrophysics, Institute of High Energy Physics, Chinese Academy of Sciences, 19B Yuquan Road, Beijing 100049, China}
\affiliation{School of Astronomy and Space Science, University of Chinese Academy of Sciences, 19A Yuquan Road, Beijing 100049, China}

\author{Yue Zhang}
\affiliation{Key Laboratory for Particle Astrophysics, Institute of High Energy Physics, Chinese Academy of Sciences, 19B Yuquan Road, Beijing 100049, China}
\affiliation{School of Astronomy and Space Science, University of Chinese Academy of Sciences, 19A Yuquan Road, Beijing 100049, China}

\author{Pu Du}
\affiliation{Key Laboratory for Particle Astrophysics, Institute of High Energy Physics, Chinese Academy of Sciences, 19B Yuquan Road, Beijing 100049, China}

\author{Yan-Rong Li}
\affiliation{Key Laboratory for Particle Astrophysics, Institute of High Energy Physics, Chinese Academy of Sciences, 19B Yuquan Road, Beijing 100049, China}

\author{Jin-Ming Bai}
\affiliation{Yunnan Observatories, Chinese Academy of Sciences, Kunming 650011, China}

\author{Luis C. Ho}
\affiliation{Kavli Institute for Astronomy and Astrophysics, Peking University, Beijing 100871, China}
\affiliation{Department of Astronomy, School of Physics, Peking University, Beijing 100871, China}

\author{Wei-Hao Bian}
\affiliation{Physics Department, Nanjing Normal University, Nanjing 210097, China}

\author{Ye-Fei Yuan}
\affiliation{Department of Astronomy, University of Science and Technology of China, Hefei 230026, China}

\author{Jian-Min Wang}
\affiliation{Key Laboratory for Particle Astrophysics, Institute of High Energy Physics, Chinese Academy of Sciences, 19B Yuquan Road, Beijing 100049, China}
\affiliation{School of Astronomy and Space Science, University of Chinese Academy of Sciences, 19A Yuquan Road, Beijing 100049, China}
\affiliation{National Astronomical Observatories of China, Chinese Academy of Sciences, 20A Datun Road, Beijing 100020, China}

%
%
%
%
%
%
%
%
%
%
%
%
%
%

\begin{abstract} 
We report results of the first reverberation mapping campaign of I Zwicky 1 during $2014$-$2016$, 
which showed unambiguous reverberations of the broad H$\beta$ line emission to the varying optical 
continuum. From analysis using several methods, we obtain a reverberation lag of 
$\tau_{\rm H\beta}=37.2^{+4.5}_{-4.9}\,$ days. Taking a virial factor of $f_{_{\rm BLR}}=1$, we find 
a black hole mass of $\mbh=9.30_{-1.38}^{+1.26}\times 10^6\sunm$ from the mean spectra. The 
accretion rate is estimated to be $203.9_{-65.8}^{+61.0}\,L_{\rm Edd}c^{-2}$, suggesting a 
super-Eddington accretor, where $L_{\rm Edd}$ is the Eddington luminosity and $c$ is the speed 
of light. By decomposing {\it Hubble Space Telescope} images, we find that the stellar mass of 
the bulge of its host galaxy is $\log (M_{\rm bulge}/\sunm) = \rm  10.92\pm 0.07$.  This leads 
to a black hole to bulge mass ratio of $\sim 10^{-4}$, which is significantly smaller than that 
of classical bulges and elliptical galaxies. After subtracting the host contamination from the 
observed luminosity, we find that I Zw 1 follows the empirical $R_{\rm BLR}\propto L_{5100}^{1/2}$ 
relation. 
\end{abstract}


\section{Introduction}

Narrow-line Seyfert 1 galaxies (NLS1s) are thought to be a special subclass of active galactic 
nuclei (AGN). Compared to the broad-line Seyfert 1 galaxies, 
NLS1s have:
(1) narrower Balmer lines (FWHM$_{\rm H\beta}\lesssim2000\kms$, by definition, 
where $\rm FWHM_{H\beta}$ is the full width at half maximum of the broad H$\beta$ emission line), 
(2) smaller intensity ratio of \oiii\ $\lambda$5007, to H$\beta$
line (\oiii/H$\beta<3$), 
(3) stronger optical \feii\, multiplet emissions,  and (4) usually 
 steeper soft X-ray spectra and more rapid X-ray variability 
(\citealt{ost85}; \citealt{good89}; \citealt{bor92}; \citealt{boller96}; 
\citealt{sulen00}; \citealt{Wang2004}). These distinctive properties of NLS1s can be explained 
by less massive black holes accreting with higher mass accretion rates 
(\citealt{bor92}; \citealt{boller96}; \citealt{wang2003}; \citealt{mathur05}; \citealt{grupe04}; \citealt{peter04}). 
Black holes in NLS1s are therefore undergoing fast growth through super-Eddington accretion (e.g.,
\citealt{Kawaguchi2004}; \citealt{Zhang2006}; \citealt{Wang2007}). 
In the high-$z$ universe, such fast growth was suggested to be a possible way of forming supermassive 
black holes 
(\citealt{Volonteri2005}; \citealt{Wang2006};  \citealt{Mortlock2011}; \citealt{Banados2018}). 
Moreover, these super-Eddington accreting massive black holes (SEAMBHs) have saturated luminosity 
predicted by slim accretion disk model \citep{Abramowicz1988,wang99} and could be used as a new kind 
of standard candle to study the expansion history of the high-$z$ 
Universe (\citealt{wang2013}; \citealt{Wang2014}; \citealt{Marziani2014}; \citealt{Cai2018};
\citealt{Marziani2019}) since they are quit common from low-$z$ to high-$z$ 
Universe \citep{du16b,Negret2018,Mary2018}. 
 Reliably measuring black hole mass of local super-Eddington AGNs 
greatly helps elucidate these issues.

 The reverberation mapping (RM) technique is a powerful tool for measuring the mass of black 
 hole in the centre of AGN \citep{Peterson1993,peter04,Peterson2014}.
  The widely accepted scenario is that 
 gas around the central back hole is photoionized by the continuum emissions from the 
 accretion disk and emits the 
 observed broad emission lines. This is known as broad-line regions (BLR).
 Because of the light-travel time from the 
 central black hole to the BLR, the variation in the flux of emission lines will delay 
 that of the continuum. RM-campaigns monitor the flux variations in the continuum and 
 the broad emission lines to measure the lags between them.  Thus the observed lags 
 are regarded as a measurement of the radius 
 of the BLR. Assuming that the motion of BLR gas is governed by the gravity of the 
 black hole, the profiles of these lines would provide kinematic information of the BLR 
 and the viral mass of the black hole can be estimated by
\begin{equation}\label{mass}
 M_{\rm \bullet}=f_{_{\rm BLR}}\frac{\it R_{_{\rm BLR}} \it V_{_{\rm {FWHM}}}^{\rm 2}}{G},
\end{equation}
where $\fblr$ is the virial factor, $G$ is the gravitational constant,
$R_{_{\rm BLR}}=c\times \tau_{\rm H\beta}$ is the emissivity-weighted  size of the 
BLR emitting the broad 
H$\beta$ line, with $c$ the light speed and $V_{_{\rm FWHM}}$ the velocity of the 
BLR clouds inferred from the width of the broad H$\beta$ line. 
 Recently, the mass of the black hole in 3C273 has been determined by another novel 
technique using GRAVITY on VLTI (Very Large Telescope Interferometer) \citep{Sturm2018}, 
and the result is  in good agreement with that given by a 10-year RM-campaign \citep{Zhang2018}.

As one of the most famous NLS1s, I Zwicky 1 (PG 0050+124 or Mrk 1502, $z=0.061$; hereafter 
I Zw 1) is selected 
as one of candidates in our large RM-campaign focusing on AGNs with SEAMBHs \citep{du14}. 
I Zw 1 is a nearby and bright ($m_{ V}=14.06 \pm 0.05$ from \citealt{sla11}) prototypical 
NLS1 (\citealt{schm83}; \citealt{ost85}; \citealt{hal87}) with 
FWHM$_{\rm H\beta} \simeq 1200 \kms$,  large relative strength of optical
\feii\, emission lines ($\mathcal{R}_{\rm Fe} = 1.47 $ from  \citealt{bor92}, 
where $\mathcal{R}_{\rm Fe}= F_{\rm Fe}/F_{\rm H\beta}$), weak intensity of \oiii\, 
line and a steep X-ray ($2-10$ keV) photon index ($\Gamma=2.15 \pm 0.02$ from \citealt{gallo07}).
Its \feii\, spectrum is commonly used as a template to fit the optical and ultraviolet spectra 
of AGNs and quasars (\citealt{bor92}; \citealt{vest01}; \citealt{vero04}). 
Similar to other NLS1s,
{\it Hubble Space Telescope} ({\it HST}) images show clearly that the host galaxy 
of I Zw 1 is a spiral galaxy with asymmetric knotty 
spiral arms (\citealt{sla11a}) consisting of young stellar populations and ongoing nuclear star 
formation activity (\citealt{hut90}; \citealt{Scharwachter2007}). 
It would be highly instructive to measure the black hole mass and accretion rate of  I Zw 1 in order to 
understand the relation of its central engine to its host galaxy. As far as we know, only \cite{gian98} 
obtained seven epochs with a mean cadence of about 170 days to monitor optical variability of I Zw 1. 
This is not enough for RM measurements.

In this paper, we report results of the first RM campaign for I Zw 1.
In \S 2, we describe the observations and data reduction. In \S 3, we analyze the light curves 
to measure the H$\beta$ time lags for black hole mass and fit the $HST$ images to separate the host 
component to test the empirical $R_{\rm BLR}$-$L_{5100}$ relation. 
Brief discussions are provided in \S4. 
A summary is given in the last section.
Throughout this work, we adopt a cosmology with $H_0 = 67\  \rm km~s^{-1}~Mpc^{-1}$, 
$\Omega_{\Lambda} = 0.68$ 
and $\Omega_{\rm m} = 0.32$ \citep{ade14} .

\section{OBSERVATIONS AND DATA REDUCTION}  
\subsection{Observations}       
\label{sect:obser}

Our observations were performed from November 2014 to February 2016 using the Lijiang 2.4 m 
telescope of Yunnan Observatories, Chinese Academy of Sciences. A versatile instrument 
named Yunnan Faint Object Spectrograph and Camera (YFOSC) was adopted since it can  switch 
from spectroscopy to photometry  within one second (\citealt{du14}).
During the spectroscopic observations, we used Grism 14, which covers the wavelength range 
of 3800$-$7200 \AA, and a long slit with a fixed width of $2.^{\pp}5$ to minimize flux loss 
caused by the different seeing conditions. The final spectral resolution was roughly 
$500 \kms$. The spectral resolution was estimated by comparing the 
width of the \oiii\  line of the Lijiang spectrum with the one obtained from the Sloan Digital Sky Survey (\citealt{hu15}; \citealt{du16}).
For accurate flux calibration, we adopted the method described by \cite{maoz90} and \cite{kas00},
putting I Zw 1 and a nearby comparison star simultaneously in the long slit 
during the exposure. The comparison star (R.A.$_{\rm J2000}$ = 00:53:46.92, 
Dec.$_{\rm J2000}$ = + 12:39:56.8) is a stable G star without intrinsic 
changes, as confirmed from the photometric light curve shown in Appendix \ref{sec:A}.
The position angle between the comparison star and I Zw 1 is $120^{\circ}$, and the angular distance is  
$3.4^{\prime}$. Flux changes caused by unstable weather can be corrected by the comparison star. 
To mitigate the influence of cosmic rays, we used two consecutive exposures with typical 
exposure time of 600 s each. Between the two exposures, we took an image to inspect whether 
I Zw 1 and the comparison star still sit in the middle of the slit. If not, we adjust the slit for realignment.
Every night we also took two consecutive exposures of a spectrophotometric standard star.
In order to minimise the impact of atmospheric differential refraction, we took spectra only 
when the air mass was $\lesssim 1.5$ (median air mass $\lesssim 1.3$ for our data). 
In addition, we also made $V$-band photometric observations to verify the accuracy of the spectral 
calibration. Three consecutive 90 s exposures were taken to reduce the impact of cosmic rays.
We obtained 66 epochs of spectroscopic observations and 74 epochs of photometric 
observations for I Zw 1 during our campaign. The median cadence is $\simeq 4$ days.

\subsection{Data reduction}

All the spectroscopic and photometric data were reduced following standard methods using IRAF v2.16. 
For spectroscopic data, the extraction aperture was $8.^{\prime\prime}5$. Standard neon 
and helium lamps were used for wavelength calibration, and the 
comparison star was used for flux calibration, as described by \cite{du14}. 
Here we briefly describe the method of flux calibration.
(1) We used spectrophotometric standard stars to calibrate the spectra of the comparison star. 
(2) A fiducial spectrum of the comparison star was made from 
its flux-calibrated spectra taken over several nights under good 
weather conditions. The uncertainty of the absolute flux calibration was $\sim 10\%$ (\citealt{du16a}).
(3) By comparing the observed spectrum of the comparison star to the calibrated fiducial spectrum, 
we obtained a sensitivity function for each exposure. 
(4) This sensitivity function was then used to calibrate the spectrum of I Zw 1.
(5) We combined the separate exposures taken each night into a single-epoch spectrum for that night.
This method can achieve an accuracy of $\sim 3\%$ in the calibrated spectra (\citealt{du18}), 
which is especially necessary for targets such as I Zw 1 and other SEAMBHs, 
whose \oiii$\lambda 5007 $ line tends to be too weak to be used for relative 
flux calibration, as conventionally practiced for RM 
(\citealt{foltz81}; \citealt {peterson1982}).

For generating the photometric light curves of I Zw 1, six stars in the same fields were used to 
perform differential photometry. The radius of the aperture for I Zw 1 was $5.^{\prime\prime}1$,
and that for the background was $8.^{\prime\prime}5 - 17.^{\prime\prime}0$. 
Specifically, for each exposure, we obtained the instrumental magnitudes of I Zw 1 and the six 
stars by IRAF. We calculated the differential magnitude between I Zw 1 and the mean of the six 
stars, and the uncertainties in all the instrumental magnitudes were propagated to the 
uncertainty in this differential magnitude. Then we averaged the differential magnitudes 
of the three exposures in the same night as the value of that individual night, and 
calculated a statistical uncertainty from the uncertainty of each exposure by error propagation. 
In addition, we calculated scatters between the differential magnitudes of the three exposures 
as the systematic uncertainty. The two uncertainties are added in quadrature as the final uncertainty 
of the differential magnitude in each night.

\section{Data analysis and results}

\subsection{Light curves}
We use the following method to determine the 5100 \AA\ flux density and the H$\beta$ flux.
The 5100 \AA\ flux is the median flux between 5075 and 5125 \AA\ in the rest frame.
The H$\beta$ flux is obtained by integrating the flux between 4810$-$4910 \AA\ from the 
continuum-subtracted spectrum, while the continuum underlying the H$\beta$ line is determined 
by a linear interpolation between two continuum bands (4740-4790 \AA\ and 5075-5125 \AA).
The continuum windows are selected to minimize the contamination from other emission 
lines, such as \oiii, \feii, and \Heii. 
In Figure \ref{meanrms}, we mark the window for the H$\beta$ line by the red band and the windows 
for the continuum by the grey bands. The measurement errors of the light curves come from both Poisson 
noise and systematic uncertainty. The systematic uncertainty arises from poor weather conditions, 
slit positioning, telescope tracking, and choice of continuum window. It can be estimated
from the standard deviation of the residuals after subtracting a median-filtered, smoothed light curve. 
Table \ref{Tab:flux} lists  the continuum flux density at 5100 \AA\ and the H$\beta$ flux. 
Only 27 epochs are included from the 2014--2015 season, while 39 epochs are available for the 
2015--2016 season. The seasonal gap between March 2015 and June 2015 is 120 days.
We show the 5100 \AA\ and H$\beta$ light curves during 2014--2016 in Figures \ref{Fig:f5100}i
and \ref{Fig:f5100}j, respectively. The large errors of some points from 2015 June 
20 to 2015 July 6 are caused by the bad weather during
the rainy season of the Lijiang Station.

To improve the sampling of the continuum light curves,  we
{ used the $V$-band photometry data from ASAS-SN (All-Sky Automated Survey for SuperNovae)\footnote{\tt{http://www.astronomy.ohio-state.edu/~assassin/index.shtml}}. 
ASAS-SN is a long-term project aiming to monitor the entire visible sky to a depth of $V < 17$ mag with a cadence of 2-3 days \citep{Shappee2014,Kochanek2017}. 
Between May 2014 and March 2016, ASAS-SN monitored I Zw 1 by two units of telescopes: ”Brutus”, deployed at the Hawaii station of the Las Cumbres Observatory, and ”Cassius”, deployed in Chile.
 73 epochs for the 2014--2015 season, 
 and 72 epochs for the 2015--2016 season, with enough S$/$N, were adopted here.}

We merge the $V$-band photometry data from 
the Lijiang and ASAS-SN into the 5100 \AA\  
light curves by applying a multiplicative scale factor and an additive flux adjustment, which 
are determined by a Markov-chain Monte Carlo (MCMC) implementation (\citealt{li14}; \citealt{du15}). 
The combined continuum light curves are shown in Figure 
\ref{Fig:fcombine}i, which contain 136 epochs from the 2014--2015 season and 149 epochs from the 
2015--2016 season. 
{
By combining ASAS-SN data, the continuum light curve shows an obvious structure for 2014-2015 season (shown in Figure \ref{Fig:fcombine}i), which corresponds to that in the light curve of H$\beta$} and the seasonal gap is shortened to 94 days. 
However, as shown in Figure \ref{Fig:fcombine}i, the 5100 \AA\ light curves increase 
the scatter of the continuum light curve, so we merge only the $V$-band photometry data from 
ASAS-SN into the photometry data from Lijiang to obtain the final photometry continuum light curve  
shown in Figure \ref{Fig:photo}i.  The final photometric continuum light curve contains
109 and 110 epochs from the 2014--2015 and 2015--2016 seasons, respectively. We adopt 
the photometric continuum light curve to obtain the H$\beta$ lags in this paper. However, for 
comparison, we also show the results 
for the 5100 \AA\ and the combined continuum light curves in Appendix \ref{sec:B}.

\subsection{Variability characteristics}
  
We use the methods described by ~\cite{rod97} to calculate light curve characteristics of the 
continuum and the H$\beta$ emission line. The variability characteristics are given by 

\begin{equation}
F_{\rm var}=\frac{(\sigma^2-\triangle^2)^{1/2}}{\langle F \rangle},
\end{equation}

where 
\begin{equation}
\sigma^2=\sum_{i=1}^{N} \frac{(F_i-\langle F \rangle)^2}{(N-1)},\quad
\triangle^2=\sum_{i=1}^{N} \frac{\Delta_i^2}{N},\quad
\langle F\rangle=\sum_{i=1}^N\frac{F_i}{N},
\end{equation}
and the error of $F_{\rm var}$ is given by
\begin{equation}
\sigma_{\rm var}=\frac{1}{F_{\rm var}} \left(\frac{1}{2 \times N}\right)^{1/2}\frac{\sigma^2}{{\langle F \rangle}^2},
\end{equation}
\citep{Edelson2002}, where
$N$ is the total number of data, $F_i$ is the flux of the $i$-th observation, and $\Delta_i$ is the 
uncertainty of $F_i$. We use $F_{\rm var}^{5100}$, $F_{\rm var}^{\rm combine}$, $F_{\rm var}^{V}$  and 
$F_{\rm var}^{\rm H \beta}$ to denote the amplitude of the variation in the light curves of the 5100 \AA\ continuum, 
the combined continuum, the photometry continuum and broad H$\beta$ line. Table \ref{Tab:var} lists $F_{\rm var}^{5100}$, $F_{\rm var}^{\rm combine}$, $F_{\rm var}^{V}$
and $F_{\rm var}^{\rm H \beta}$ for the 2014--2015 data, 2015--2016 data, and 2014--2016 data, respectively.
We find that $F_{\rm var}^{5100}$ , $F_{\rm var}^{\rm combine}$, $F_{\rm var}^{V}$ and 
$F_{\rm var}^{\rm H \beta}$ of I Zw 1 are clearly much smaller than that of other PG quasars (see Table 
5 in \citealt{peter98} and Table 5 in \citealt{kas00}), by a factor of a few.

\begin{figure*}[t!]
\begin{center}
\includegraphics[angle=0, width=0.8\textwidth]{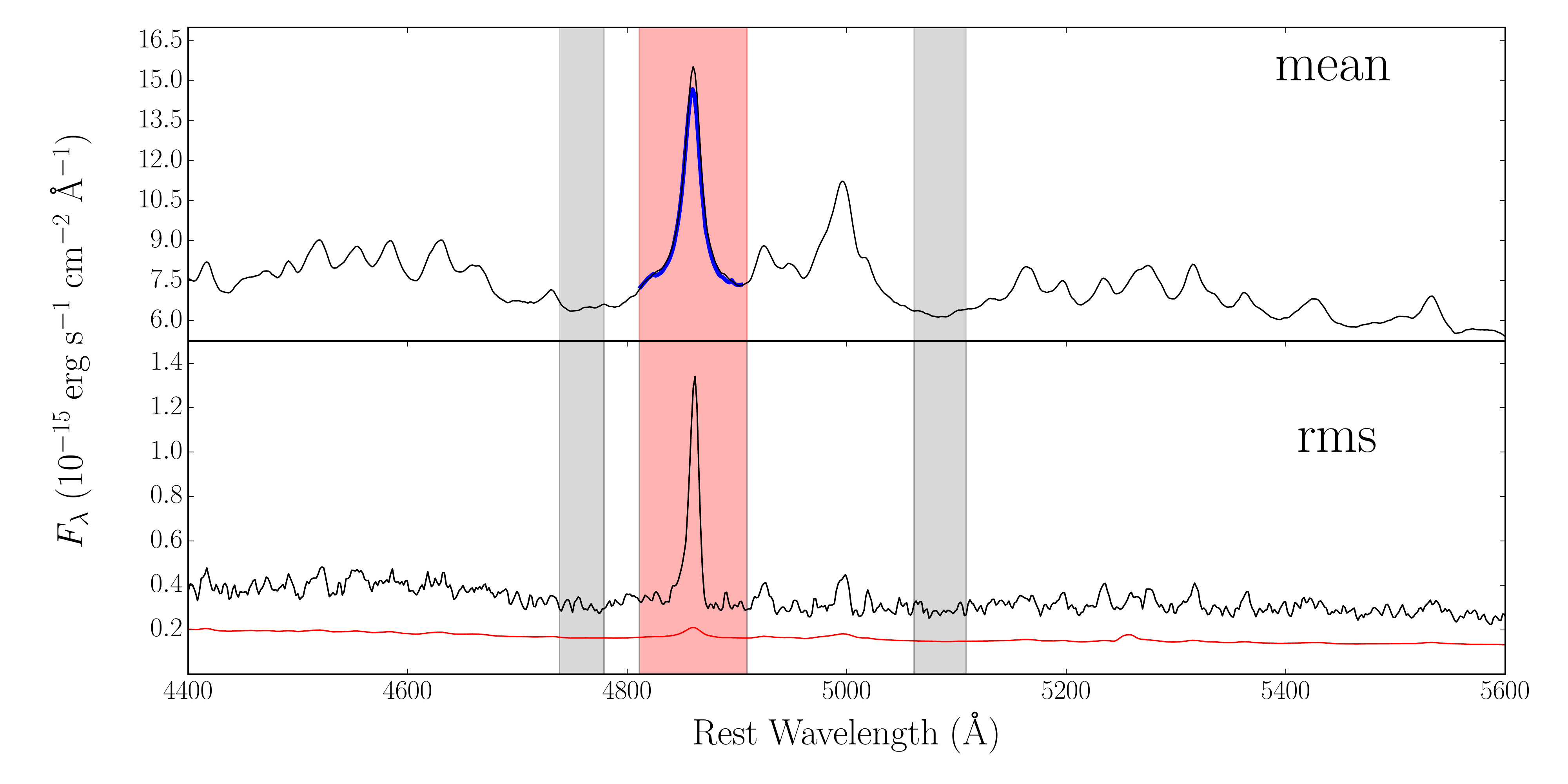}
\end{center}
\vglue -0.5cm
\caption{\footnotesize 
The mean and rms spectra for I Zw 1 from the 2014--2015 season. The grey bands are the continuum windows 
and the red band is the integration window for H$\beta$. The blue line is the narrow line-subtracted profile, 
and the red line is the mean of the errors of the individual spectra.}
\label{meanrms}
\end{figure*}

The mean and the rms spectra of I Zw 1 are calculated by
\begin{equation}
\bar{F}({\lambda})=\frac{1}{N}\sum_{i=1}^{N}F_{i}(\lambda);\quad
S({\lambda})=\left\lbrace                                                                                                                                                                                                                                                                                                                                                                                                                                                                                                                                                                                                                                                                                                                                                                                                                                                                                                                                                                                                                                                                                                                                                                                                                                                                                                                                                                                                                                                                                                                                                                                                                                                                                                                                                                                                                                                                                                                                                                                                                                                                                                                                        \frac{1}{N-1}\sum_{i=1}^N \left[F_{i}(\lambda)-\bar{F}(\lambda) \right]^2 \right\rbrace^{1/2}.
\end{equation}
We show the results for the 2014--2015 season in Figure \ref{meanrms}. The \oiii\, line disappears in the 
rms spectrum, which implies that the spectral calibration is quite good. 
We also plot the mean of the error of the individual spectra (see the red line in Figure \ref{meanrms}), 
which can be taken as the rms spectrum for the variance caused by noise only (\citealt{hu16}).

\subsection{$H\beta$ lags}
We employ three methods to calculate H$\beta$ lags: the interpolation 
cross-correlation function (ICCF; \citealt{gas86}; \citealt{peter04}), JAVELIN (\citealt{zu11}), and 
MICA\footnote{MICA 
is available at {\tt https://github.com/LiyrAstroph/MICA2}.} (\citealt{li16}). The ICCF method calculates the cross-correlation function (CCF) between the light 
curves of the continuum and H$\beta$ fluxes. The time lag is determined by measuring either the location  
$\tau_{\rm peak} $ of the CCF peak ($r_{\rm max}$) or the centroid $\tau_{\rm cent} $ of the points 
around the peak above a threshold $r \geq 0.8\ r_{\rm max}$. The associated uncertainties are estimated 
by the Monte Carlo flux randomization/random subset sampling (FR/RSS) method of \cite{peter98} and are 
given at 63.8$\%$ confidence levels. Both JAVELIN and MICA are forward-modelling methods that use the damped random 
walk model (e.g., \citealt{kel09}; \citealt{zu11}; \citealt{li13}) to delineate the variations of continuum 
light curves and directly infer the transfer functions of the BLR by presuming specific shapes of the transfer
functions. JAVELIN adopts a top-hat transfer function, whereas MICA expresses the transfer function as a sum of 
a series of Gaussians. For simplicity, we only use one Gaussian in MICA. Accordingly, time lags are given by 
the center of the top-hat function in JAVELIN and the center of the Gaussian in MICA. Both JAVELIN and MICA 
use the MCMC technique to explore the model parameters of transfer functions. The time lags and the associated
uncertainties are estimated as the expectation and standard deviation of the generated Markov chains, 
respectively.

\begin{figure*}[t!]
\begin{center}
\includegraphics[angle=0, width=\textwidth]{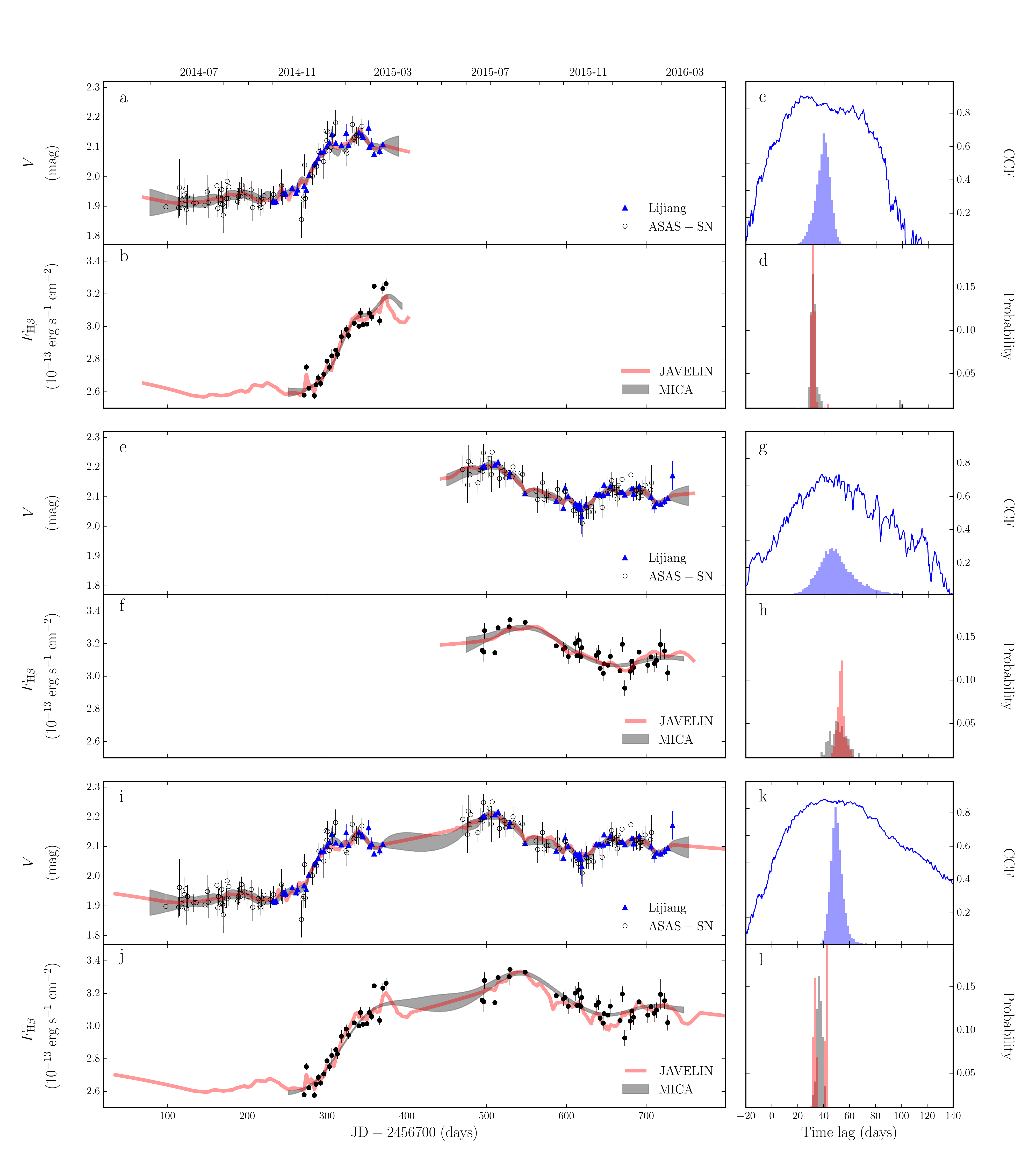}
\end{center}
\caption{\footnotesize 
Light curves and results of correlation analysis using the photometric continuum light curves for I Zw 1. 
The {\it left} panels are 
light curves for different seasons. The red lines are the continuum light curve fitted by 
JAVELIN, while the gray bands are those fitted by MICA. The {\it right} panels show the correlation 
analysis. Panels ({\it c, g, k}) show ICCF results and panels ({\it d, h, l}) show results for MICA and 
JAVELIN for season 2014-2015, 2015-2016 and 2014-2016, respectively.
The blue lines are the cross-correlation 
functions and the blue histograms are the cross correlation centroid distributions. 
The red histograms show the 
posterior probability of the time lags for JAVELIN, and the gray histograms show the corresponding one 
for MICA. JAVELIN, MICA and ICCF produce consistent
lags for the same data. 
}
\label{Fig:photo}
\end{figure*}

In order to avoid potential biases introduced by seasonal gaps, we 
analyze the observations from the 2014--2015 and 2015--2016 seasons, as well 
as that for the whole campaign, separately (Figure \ref{Fig:photo}).  Figures 
\ref{Fig:photo}a and \ref{Fig:photo}b show the light curves for the photometric continuum
and H$\beta$ emission line for the season 2014--2015.  The light curves 
reconstructed by JAVELIN (red line) and MICA (gray band) are also superposed. 
Figure \ref{Fig:photo}c shows the CCF and the cross-correlation centroid 
distribution (CCCD) for the season 2014--2015, and Figure \ref{Fig:photo}d
gives the posterior distributions of time lags obtained by JAVELIN (red) and 
MICA (gray).  Figures \ref{Fig:photo}e--\ref{Fig:photo}h are the same as Figures
\ref{Fig:photo}a--\ref{Fig:photo}d, but for the season 2015--2016, and the results for 
the entire 2014--2016 season are shown in Figures \ref{Fig:photo}i--\ref{Fig:photo}l.
Table \ref{Tab:lags} summarises the time lags determined by the above three methods. They 
are consistent to each other, within uncertainties.

We detect significant time lags with small uncertainties using the photometric 
light curves.  The H$\beta$ time lags for the 2015--2016 data have large 
uncertainties (Figures \ref{Fig:photo}g and \ref{Fig:photo}h) because of the small 
variability amplitude and large flux dispersion of the light curves (see 
Figures \ref{Fig:photo}e and \ref{Fig:photo}f).  In principle, JAVELIN and MICA can 
reconstruct the light curves during the large seasonal gaps, such that, 
for data for the whole campaign, the lags obtained by JAVELIN and MICA should 
be more robust than that derived by ICCF.  We also find that the lags obtained 
using JAVELIN and MICA (Figure \ref{Fig:photo}l) are more consistent with 
that obtained by ICCF for the 2014--2015 data (Figure \ref{Fig:photo}c).
From inspection of the CCF peak value $r_{\rm max}$, we adopt 
$\tau_{\rm cent}=37.2^{+4.5}_{-4.9}$ days from the ICCF analysis of 
the photometric continuum and H$\beta$ light curves from 2014$-$2015, 
and we use this value to calculate the black hole mass below.

\subsection{Host galaxy}

We use images obtained with the {\it HST}\ to study the host galaxy, in particular its 
overall morphology, bulge-to-total ratio, and bulge stellar mass.  We also need to 
determine the degree to which the spectroscopically measured optical continuum is 
contaminated by host galaxy emission.  The highest quality {\it HST}\ observations 
currently available are those obtained with the WFC3 camera on 2 November 2013 
(GO-12903, PI: Luis C. Ho).  
I Zw 1 was observed for 300 s with the F438W filter (395--468 nm) in the UVIS channel 
and for 147 s with the F105W filter (901--1204 nm) in the IR channel.  An additional 
short (40 s) F438W exposure was taken to warrant against saturation of the nucleus in 
the long exposure.  These two filters were purposefully chosen to mimic $B$ and $I$ 
in the rest frame of I Zw 1, at the same time avoiding contamination from strong 
emission lines.  The large wavelength separation between the two filters also offers 
the most leverage for constraining the stellar population (Section 5.2).  To better 
sample the point-spread function (PSF), the long UVIS observation was taken with a 
three-point linear dither 
pattern, while the IR observation was taken with the four-point boxy dither pattern.  
To avoid overheads due to buffer dump, we employed the UVIS2-M1K1C-SUB 1k$\times$1k 
subarray for the UVIS channel and the IRSUB512 subarray for the IR channel, which 
resulted in a restricted field-of-view of 40$^{\prime\prime}$ $\times$40$^{\prime\prime}$ 
and 67$^{\prime\prime}$ $\times$67$^{\prime\prime}$, respectively.  

Because of the sparseness of field stars in the vicinity of I Zw 1, the observations 
were conducted in GYRO mode, which, unfortunately, led to considerable degradation of 
the PSF of the dither-combined image generated from the standard 
data reduction pipeline.  Instead, we use the DrizzlePac task AstroDrizzle (v1.1.16) to 
correct the geometric distortion, align the sub-exposures, perform sky subtraction, 
remove cosmic rays, and, finally combine the different exposures.  The core of the AGN 
was severely saturated in the long F438W exposure, and it was replaced with an 
appropriately scaled version of the 40 s short exposure. Because the PSF of 
{\it HST} is undersampled, we broaden both our science and PSF images by 
convolving them with a Gaussian kernel so that the PSF 
can reach Nyquist sampling \citep{Kim2008a}. This largely removes the sub-pixel mismatch 
of the core of the PSF and the smearing of the nucleus due to image drift.

\begin{figure*}[t!]
\begin{center}
\includegraphics[angle=0, width=\textwidth]{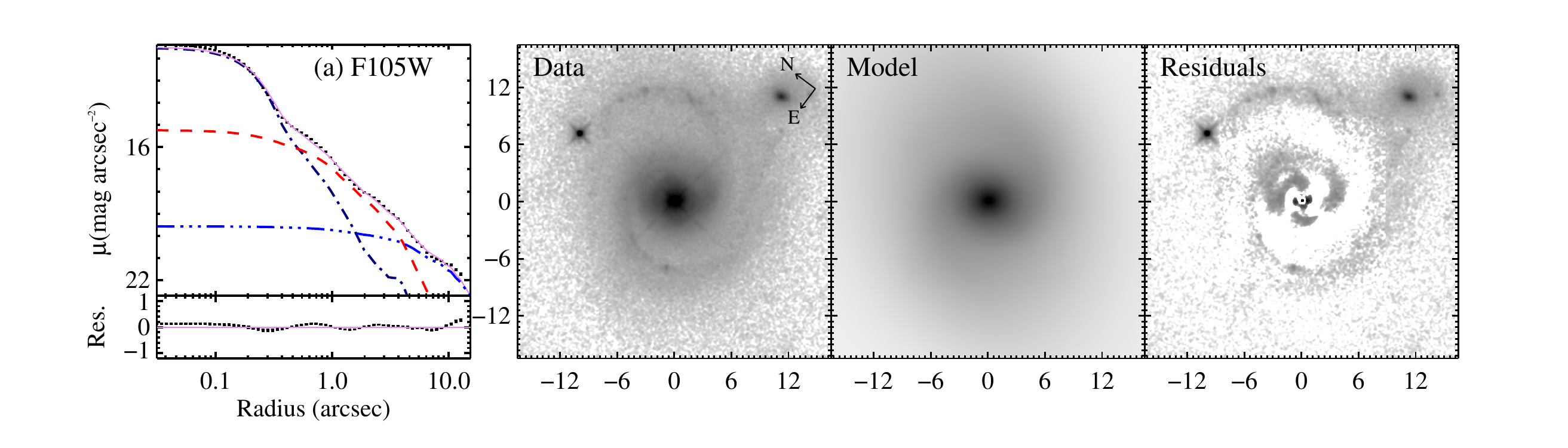}
\includegraphics[angle=0, width=\textwidth]{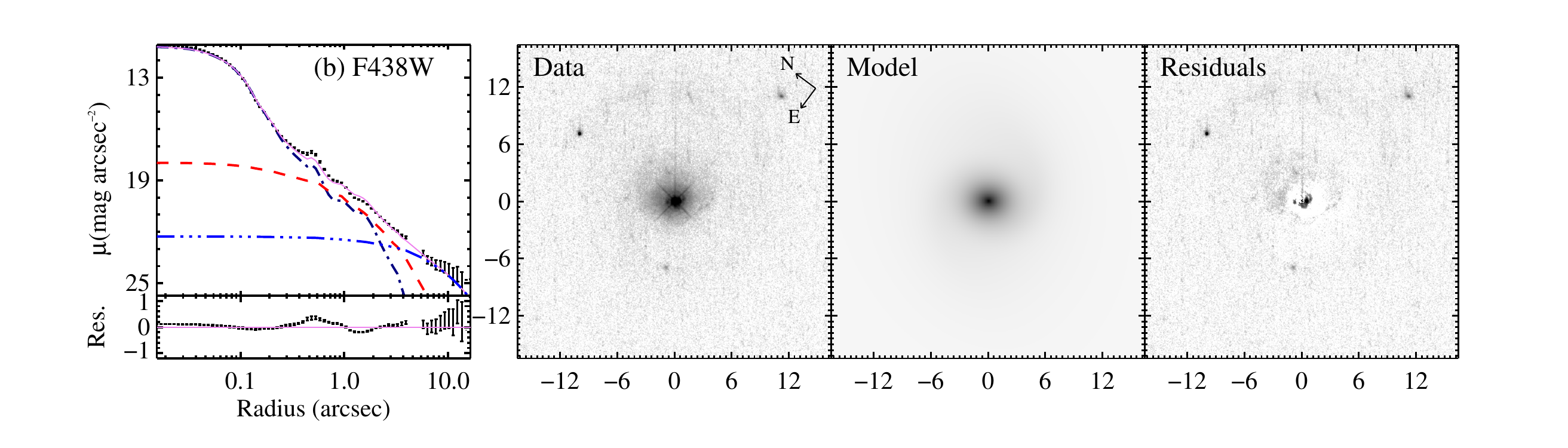}
\end{center}
\vglue -0.5cm
\caption{\footnotesize 
GALFIT decomposition for I Zw 1. The upper and lower panels show the fitting result of two different 
filters. In each row, 1D profile, 2D image of original data, best-fit model for the host (the AGN is 
excluded to better highlight the host) and residuals images are displayed. The 1D surface brightness 
profiles show the original data (black dots), the best fit (violet solid line), and the subcomponents 
(PSF in dark blue dot-dashed line, bulge in red dashed line and disk in blue dot-dashed line). The 
units of the images are in arcseconds. All images are on an asinh stretch. }
\label{hst}
\end{figure*}

The F105W image (Figure \ref{hst}a) reveals a nearly face-on spiral galaxy with 
two dominant spiral arms.  A bar-like structure may also be present.  The F438W 
image (Figure \ref{hst}b) is considerably shallower, but the host is still clearly 
detected.  To extract quantitative measurements of the bulge, we use the program  
GALFIT (v3.0.5; \citealt{Peng2002, Peng2010} ) to fit
two-dimensional surface brightness distributions to the {\it HST} images.  A crucial 
ingredient is the PSF, which will have a strong effect on the bright and active nucleus.  
Unfortunately, no suitably bright star is available to be used as the PSF within the 
limited field-of-view of the subarray WFC3 images.  Instead, we generated a high-S/N 
stacked empirical PSF by combining a large number (24 for F105W and 57 for F438W) of 
bright, isolated, unsaturated stars observed during the course of other WFC3 programs. 
Extensive tests, consisting of fits to isolated bright stars, indicate that 
our stacked empirical PSF is far superior to synthetic PSFs generated from the TinyTim 
program \citep{Krist1999}, and it has higher S/N than the PSFs of individual stars. 
The reduced $\chi^2$ of the fits are $\sim$3 times larger for the TinyTim synthetic PSF.  
Comparison of empirical PSFs observed from different programs indicate that the WFC3 
PSF does not vary significantly with time ($<$ 10\%).

We concentrate first on obtaining the best global fit on the deeper F105W image, whose 
red wavelength is also more sensitive to the host.  After much experimentation, we adopt 
a model with three components: a point source (represented by the PSF) for the nucleus, 
a bulge parameterized as a \cite{sersic1968} function with index $n$, and an exponential 
disk (equivalent to a S\'ersic function with $n = 1$), with coordinate rotation turned 
on to fit the spiral arms.  We could not obtain a robust solution that considers the 
faint, bar-like feature, and in the end we did not treat it.  We use the $m=1$  Fourier 
mode of the disk, which is sensitive to lopsidedness, to gauge the degree of global 
asymmetry of the galaxy (see, e.g., \citealt{Kim2008b,Kim2017}). The best model 
(Figure \ref{hst}a) reveals a bulge with $n = 1.73\pm0.06$, formally but barely 
below the conventional threshold for pseudo-bulges ($n < 2$; \citealt{Fisher2008}) and 
an overall $B/T = 0.52 \pm {0.04}$.  The disk is moderately asymmetric, with Fourier 
amplitude $a_1 = 0.11 \pm {0.01}$.  As the host is considerably weaker in the 
F438W image, we fit it keeping the structural parameters fixed to the values obtained 
from the F105W model, solving only for the magnitude.  

The uncertainties of the decomposition are dominated by PSF 
mismatch in the nucleus. We estimate the effect of the PSF by generating variants of 
the empirical PSF by combining different subsets of stars, and then repeating the fit. 
The final error budget is the quadrature sum of these two contributions.

\subsection{The $R-L$ relation}
A proper evaluation of the $R-L$ relation must consider the influence of host galaxy 
contamination on the luminosity (\citealt{kas00}; \citealt{bentz13}).  We use the 
decomposition of the {\it HST}\ images of I Zw 1 to estimate the contribution of the host 
galaxy within the spectral extraction aperture.  The host flux at 5100 \AA\ is transformed 
from the F438W magnitude with the IRAF task {\tt synphot}, assuming a stellar population 
template with an age of 5.0 Gyr (Section 4.2), after correcting for redshift
($z=0.0611$) and Galactic reddening [$E(B-V)=0.057$ mag; \citealt{schlafly11}] using 
the extinction curve of \cite{cardelli1989}.  

Table \ref{Tab:para} lists the observed total, host galaxy, and AGN fluxes at 5100 \AA.  With an AGN 
luminosity of $L_ {5100} = 3.19\times 10^{44}\ergs$, I Zw 1 follows the empirical 
$R_{\rm BLR}-L_{5100}$ relation (Figure \ref{rl}a).

\section{Black hole mass and accretion rate}
\subsection{Mass}

 For measuring the black hole mass, the velocity of the BLR clouds are measured through the FWHM of the broad emission line (Equation \ref{mass}). The narrow H$\beta$ component ($F_{\rm N}$) of I Zw 1 is too weak to be decomposed directly by spectral fitting even for the mean spectrum. Thus we subtract the $F_{\rm N}$  
from the mean spectrum of the 2014--2015 season assuming
$F_{\rm N}/F_{[\rm O\ {III}]} \simeq 0.1$  
\citep{hu12}, where $F_{[\rm O\ {III}]}$ is the flux of 
\oiii $\ \lambda 5007\ $\AA.  Then we measure the FWHM directly from the narrow line-subtracted 
profile (blue line in Figure \ref{meanrms}). The uncertainty is bracketed by assuming that 
$F_{\rm N}/F_{[\rm O\ {III}]} = 0$ and 0.2. Correcting for an 
instrumental broadening of $500 \kms$ (\citealt{hu15}; \citealt{du16}), 
$V_{_{\rm FWHM}} =\rm FWHM_{\rm H\beta}=1131_{-38}^{+35}\,\kms$ (Table 4). 
Adopting $\tau_{\rm H\beta} = 37.2$ days and $\fblr = 1$, Equation (\ref{mass}) yields
$\mbh=9.30_{-1.38}^{+1.26}\times 10^6\,\sunm$. It is known that the observed 
kinematics of the BLR are generally influenced by inclination effects, which will lead to 
uncertainties of the virial factor (\citealt{Krolik2001}; \citealt{collin06}). In principle, 
$\fblr$ can be calibrated using the $\mbh-\sigma$ relation of inactive galaxies, and the 
current best estimates of $\fblr$ are $1.3 \pm 0.4$ for classical bulges and $0.5 \pm 0.2$ 
for pseudo bulges, when $\fblr$ is calibrated using FWHM based on mean spectra 
(\citealt{ho14}).  The host galaxy of I Zw 1 likely contains a pseudo bulge (Section 5.2).  
However, the large scatter of the $\mbh-\sigma$ relation of pseudo bulges (Kormendy \& Ho 
2013) introduces significant uncertainty into the estimate of $\fblr$.  For consistency 
with our previous work on SEAMBHs \citep{du14, du15, du16}, we adopt $\fblr=1.0$ 
(\citealt{netzer10}; \citealt{woo13}).
We also estimate the black hole mass from the FWHM of the rms spectrum 
($	V_{_{\rm FWHM}}^{\rm rms} =\rm FWHM_{\rm H\beta}^{\rm rms} = 606^{+28}_{-28}\,\kms$) using $\fblr^{\rm rms}=1.12$ 
(\citealt{woo15}) and obtain $\mbh^{\rm rms}=2.99_{-0.48}^{+0.46}\times 10^6\,\sunm$.

Meanwhile, polarized spectra of type 1 AGNs can provide invaluable information for
estimating black hole masses.  Considering electron scattering in the equatorial plane 
(\citealt{Smith2005}), a polarized spectrum is equivalent to a spectrum seen by an observer 
on the mid-plane.  Thus, polarized spectra can yield more reliable estimates of the black 
hole mass (\citealt{Afanasiev2015}; \citealt{baldi16}; \citealt{Songsheng2018}). 
For equatorial scattering of a BLR with a flattened geometry, 
\begin{equation}
\label{polarize}
\mbh^{\prime}/\sunm=f_{_{\rm BLR}}^{\prime}\frac{R_{_{\rm BLR}} V_{_{\rm FWHM}}^{\prime^2}}{G}
                =8.18\times 10^6\,\left(\frac{\tau_{\rm H\beta}}{37.2\rm \ days}\right)
                \left(\frac{V_{_{\rm FWHM}}^{\prime}}{2165\kms}\right)^2,
\end{equation}
where the virial factor $\fblr^{\prime}\approx 0.24$ (with small uncertainty) 
and $V_{_{\rm FWHM}}^{\prime}$ is the FWHM of the broad-line profile in the polarized 
spectrum (Figure 3 in \citealt{Songsheng2018}).   Fortunately, a polarized spectrum of 
I Zw 1 for the H$\alpha$ emission line has been taken using the 6 m telescope of the Special 
Astrophysical Observatory of the Russian Academy of Science (L. Popovi\'c 2018, private 
communications)\footnote{{\tt https://zenodo.org/record/1219726\#.WxDfUO6FPIU}}.  We
find FWHM$_{\rm H\alpha}^{\prime}=1983\kms$ from the polarized spectrum.
 The relation between the widths of H$\beta$ and H$\alpha$ in the polarized spectrum is unclear, but a reasonable assumption is that the same relation is followed as in the total spectrum if 1) both lines are scattered by the same population of electrons, and 2) the regions emitting H$\beta$ and H$\alpha$ lines are much smaller than the electron scattering region. So we use the relation FWHM$_{\rm H\beta}=1.07\times 10^3\left({\rm FWHM_{H\alpha}}/10^3\kms\right)^{1.03}$ from \cite{Greene2005}, and obtained $V_{_{\rm{ FWHM}}}^{\prime}=\rm {FWHM}_{\rm H\beta}^{\prime} = 2165_{-182}^{+189}\kms$. This value needs to be tested observationally in the future.
Equation (\ref{polarize}) then yields $\mbh^{\prime}=8.18_{-1.75}^{+1.74}\times 10^6\sunm$, which agrees 
remarkably well with the mass estimates based on the mean spectrum. 

\subsection{Accretion rates}
Photon trapping causes the radiated luminosity of slim disks to saturate 
(\citealt{Abramowicz1988}; \citealt{wang99}; \citealt{Mineshige2000}). Under these conditions,
the normal Eddington ratio $L_{\rm bol}/L_{\rm Edd}$ is very insensitive to the accretion 
rate and mainly depends on the black hole mass, where $L_{\rm Edd}$ is the Eddington 
luminosity and $L_{\rm bol}$ is the bolometric luminosity.  This renders $L_{\rm bol}/L_{\rm Edd}$ 
unsuitable to indicate the accretion rate of SEAMBHs. As derived from the self-similar solution of 
slim disk, the photon trapping radius is $R_{\rm trap}\approx 72\,(\mathdotM/80)R_{\rm g}$, where 
$R_{\rm g}=G\mbh/c^2$, and the radius emitting optical (5100 \AA\ ) photons is 
$R_{\rm 5100} \approx 4.3 \times 10^{3} m_{7}^{-1/2} R_{\rm g}$ 
(\citealt{wang99}), which is much larger than $R_{\rm trap}$. So optical 
photons can escape freely, and hence accretion rates can be reliably estimated from 
the formalism of the standard accretion disk model \citep{Shakura1973}. The dimensionless 
accretion rate is defined as $\mathdotM=\dot{M}_{\bullet}/L_{\rm Edd}c^{-2}$, with
$\dot{M}_{\bullet}$ is the mass accretion rate. Given the 5100 \AA\ luminosity and the 
black hole mass, $\mathdotM=20.1(\ell_{44}/\cos i)^{3/2}m_7^{-2}$, where
$\ell_{44}=L_{5100}/10^{44} \ergs$, $m_7 = M_{\bullet} /10^7 M_{\odot}$, and $i$ is the 
inclination angle between the line-of-sight and the accretion disk \citep{du15}.  For I Zw 1,
we have
\begin{equation}
\mathdotM\approx  203.9\,\left(\frac{\ell_{44}}{3.19}\right)^{3/2}
                         \left(\frac{m_7}{0.93}\right)^{-2},
\end{equation}
yielding $\mathdotM=203.9_{-65.8}^{+61.0}$, for $\rm cos \it i$ = 0.75.  
As discussed in \cite{du16}, $\rm cos \it i$ = 0.75 is a reasonable mean 
value for type 1 AGNs.  If we take 
$\Delta \rm \  log \ cos\  \it i \lesssim \rm 0.1$, then the uncertainty on $\mathdotM$ 
due to $i$ will be $\Delta\  \rm log\ \mathdotM = 1.5\  \Delta \ log \ cos\  \it i \lesssim \rm 0.15$. 
 { The inclination also has insignificant effect on the observed width of the broad emission lines $V_{_{\rm {FWHM}}}$ . If we know the Keplerian velocity $V_{_{\rm K}}$ and the height $H_{\rm BLR}$ of the flattened BLR,  by the zeroth-order approximation, $
V_{_{\rm FWHM}} \approx \left[ \left({H_{\rm BLR}}/{R} \right)^2 +  \sin^2  i\right]^{\rm 1/2}\ V_{_{\rm K}}$. By detailed modeling of RM data, \cite{li13} and \cite{pan14} suggest ${H_{\rm BLR}/R} \sim 1$, thus ${\rm sin \it i} \le {H_{\rm BLR}/R} $,  which implies the inclination has insignificant influence on $V_{_{\rm {FWHM}}}$, hence  $M_{\bullet}$ and then ${\dot{\mathscr{M}}}$.} 
The high dimensionless accretion rate indicates that I Zw 1 is a SEAMBH. 
 Note that the accretion rate could be even higher by a factor of $\sim$ 10 than the present, if the FWHM obtained from rms spectrum are used in the calculation.

\section{Discussion}

\subsection{The $R-L$ relation}
\begin{figure*}[]
\begin{center}
\includegraphics[angle=0, width=0.45\textwidth]{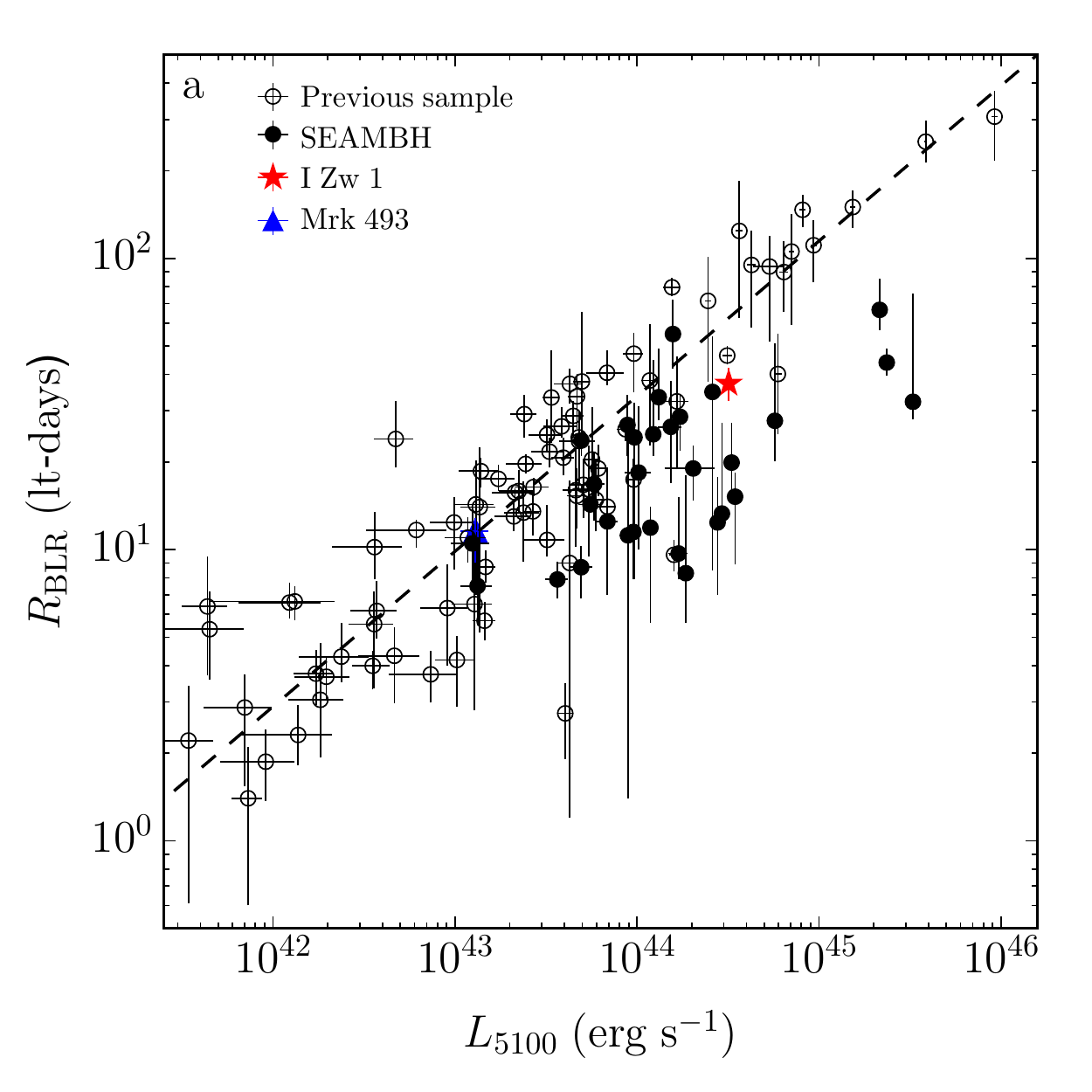}
\includegraphics[angle=0, width=0.45\textwidth]{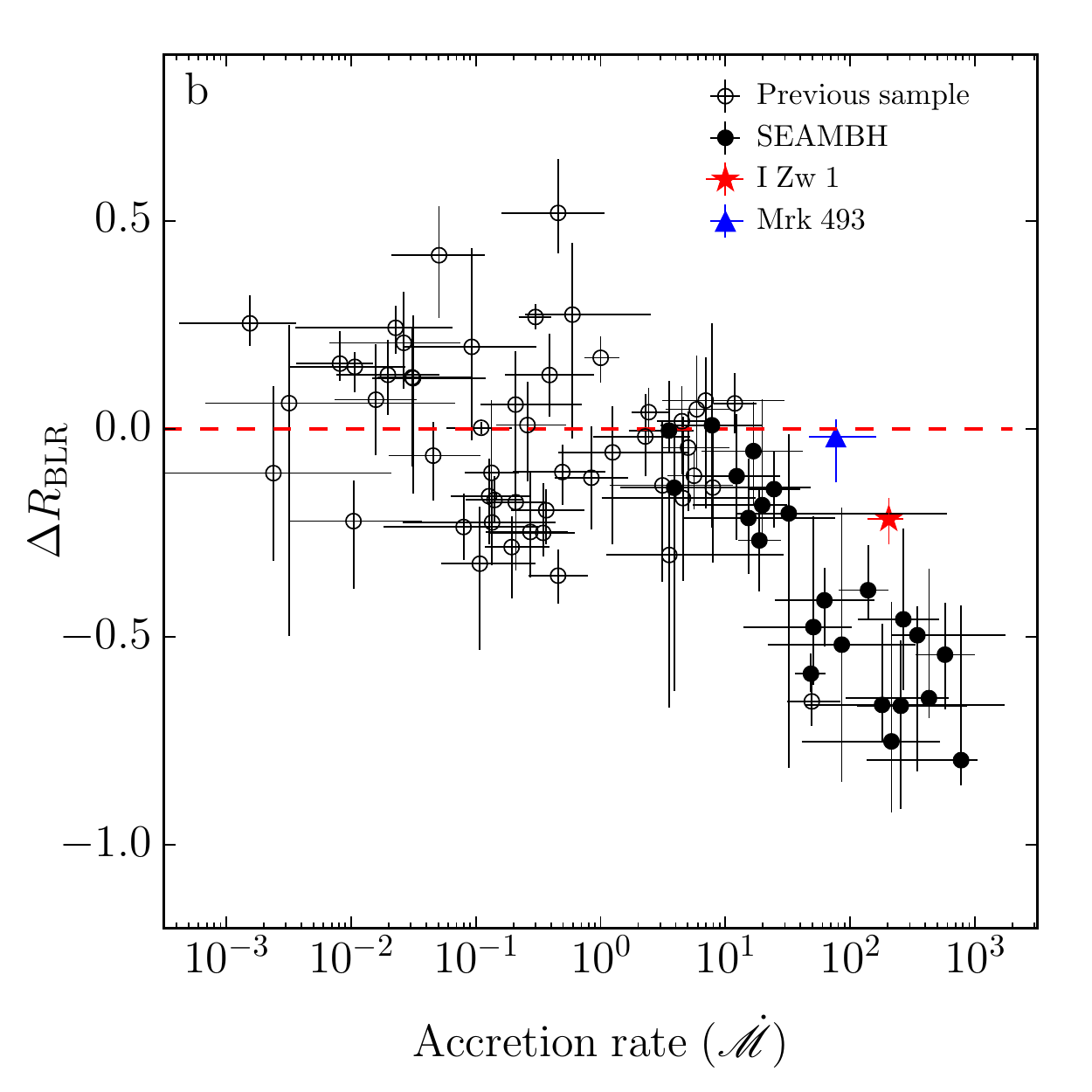}
\end{center}
\vglue -0.5cm
\caption{\footnotesize 
The (a) $R_{\rm BLR}-L_{5100}$ and  (b) $\Delta R_{\rm BLR}-\dot{\mathscr{M}}$ plots, in which  
the open and solid points are from \cite{bentz13} and \cite{du18}, respectively. 
The black dashed line is the classical $R-L$ relation \citep{bentz13}.
The red star is I Zw 1, and the blue triangle is Mrk 493 from \cite{hu15}; both are  outliers in panel {\it b}.
In panel {\it b}, the red dashed line is $\Delta R_{\rm BLR}=0$. I Zw 1 and Mrk 493 have
$(\Delta R_{\rm BLR}, \log \mathdotM)=(-0.22,2.31), (-0.02,1.88)$, respectively. }
\label{rl}
\end{figure*}

There is growing evidence that the empirical $R_{\rm BLR}$-$L_{5100}$ relation does not 
apply to SEAMBHs. For a given luminosity, SEAMBHs tend to have a shorter H$\beta$ lag,
suggesting that the size of the BLR is related not only to luminosity but also to the 
accretion rate \citep{du15,du16,du18}. According to the unified scaling relation for sub- 
and super-Eddington AGNs suggested by \cite{du16},
\begin{equation}\label{rlall}
R_{\rm BLR}=29.6\,\ell_{44}^{0.56} \min\left[1,\left(\frac{\mathdotM}{11.2}\right)^{-0.52}\right] 
\ {\rm lt\verb|-|day},
\end{equation}
I Zw 1 is predicted to have an H$\beta$ delay of 12.5 days, which is much shorter 
than our measured value of 37.2 days. Unlike other SEAMBHs, I Zw 1 actually follows the 
standard $R_{\rm BLR}$-$L_{5100}$ relation and does not show an obvious shortened H$\beta$ 
lag (Figure \ref{rl}a).  Defining the deviation 
$\Delta R_{\rm BLR} \equiv \log (R_{\rm BLR}/R_{\rm R-L})$, where $R_{\rm R-L}$ is given by 
the empirical $R_{\rm BLR}-L_{5100}$ relation, Figure \ref{rl}b plots $\Delta R_{\rm BLR}$ 
versus $\mathdotM$.  I Zw 1 deviates from the trend defined by most SEAMBHs. So does Mrk 493.
The reasons for these outliers are unclear. One possibility is that some lags are 
too short to detect, given the current observation cadence.  Within the SEAMBH framework 
described by \cite{Wang2014b} and \cite{du18}, { there are two BLRs: a normal one 
unshadowed, and another shadowed by the inner part of the disk and much closer to the 
central black hole. The emissivity-weighted gas distribution yields two peaks in the 
transfer function, corresponding to the unshadowed and shadowed BLRs respectively. 
Such a two-region BLR scenario gets supported from a recent modelling of the observed 
RM data in Mrk 142 by \cite{Li2018}, which 
constructed flexible one-region and two-region models that included
the spatial distribution and anisotropic emissivity of BLR clouds. They found that 
the two-region model is preferrable to the one-region model.
However, the observed H$\beta$ lag depends on the data quality, especially the 
cadence. For example, low cadence will smooth the light curves and yields a 
single broad peak in the CCF, although there are two peaks in the transfer function.} 
To measure the potential shorter lag, campaign with higher cadence than the present 
are being planned for getting detailed BLRs of I Zw 1 and Mrk 493 as done in Mrk 142.

\subsection{Black hole and bulge stellar masses}

The GALFIT decomposition of the {\it HST}\ images of I Zw 1 yields a bulge magnitude 
$17.23 \pm 0.04$ mag in F438W and $14.07 \pm 0.07$ mag in F105W. We apply $K$-correction 
to convert the {\it HST}-based magnitudes to rest-frame magnitudes in $B$ and $I$. 
We generate a series of template spectra with ages spanning 1 to 12 Gyr, adopting a 
\cite{Bruzual2003} models with solar metallicity, a \cite{Chabrier2003} stellar initial 
mass function, and an exponentially decreasing star formation history with a star formation 
timescale of 0.6 Gyr.  After accounting for Galactic extinction and redshift, we convolve 
the spectra with the response functions of the {\it HST}\ filters and generate synthetic 
F438W and F105W magnitudes.  The observed bulge color of F438W$-$F105W = $3.16 \pm 0.08$ 
mag is best matched with a stellar population of age $5.0^{+1.0}_{-0.5}$ Gyr,
resulting in rest-frame $M_I=-22.70\pm0.07$ mag and $B-I = 1.96 \pm 0.09$
mag.  From Table 4 of \cite{Bell2001}, we derived $M_\mathrm{bulge} = 10^{11.18\pm0.07} M_\odot$, 
assuming solar metallicity and a \cite{Salpeter1955} initial mass function.  A Chabrier initial 
mass function gives 45\% less mass \citep{Longhetti2009},  
and therefore $M_\mathrm{bulge} = {10^{10.92\pm0.07}} M_\odot$.

For $M_\bullet = 10^{6.97} M_\odot$, $M_\bullet/M_\mathrm{bulge} \approx 10^{-4}$, which is lower 
by a factor of $\sim 50$ than the value inferred from the $\mbh-M_{\rm bulge}$ relation for classical 
bulges and elliptical galaxies, but lies within the large scatter and lower ratios found for pseudo 
bulges \citep{Kormendy2013}.   I Zw 1 very likely hosts a pseudo bulge, in view of its relatively low 
S\'ersic index of $n = 1.73\pm 0.06$ (\citealt{Fisher2008}) and evidence for recent (see discussion above) 
and ongoing \citep{Scharwachter2007} nuclear star formation.

\section{SUMMARY}
The first campaign for I Zw 1 had been performed by the SEAMBH project during 2014$-$2016.  
The main conclusions are as follows.
\begin{enumerate}
\item Applying the ICCF centroid method for the 2014$-$2015 data, the H$\beta$ time lag is
$\tau_{\rm H \beta} =37.2^{+4.5}_{-4.9}$ days, and it follows the empirical $R_{\rm BLR}-L_{5100}$ 
relationship.  
 
\item Using the total mean spectra, we calculate a black hole mass of
$\mbh = 9.30_{-1.38}^{+1.26}\times 10^6\, \sunm$ and an accretion rate of
$ 203.9_{-65.8}^{+61.0}\ L_{\rm Edd}c^{-2}$. 
From the rms spectrum, we estimate $\mbh= 2.99_{-0.48}^{+0.46}\times 10^6\,\sunm$,
whereas the polarized spectrum yields $\mbh=8.18_{-1.75}^{+1.74}\times 10^6\,\sunm$.

\item We derive the stellar for the bulge of the host galaxy from 
detailed decomposition of {\it HST} images.  We find 
a mass ratio of $\mbh/M_{\rm bulge}\approx 10^{-4}$, much lower than that for 
nearby inactive classical bulges and elliptical galaxies.
\end{enumerate}
 
I Zw 1 is famous for its strong and narrow \feii\ lines; however, the current 
data do not allow us to successfully measure a lag for \feii.  We are in 
the process of scheduling a new monitoring campaign with improved cadence and 
homogeneity in observation more favourable for detecting \feii\ lags.

\acknowledgements{ 
The authors are grateful to an anonymous referee for helpful reports on this paper.
This research
is supported by National Key Program for Science and Technology Research and Development(grant
2016YFA0400701 and grants NSFC-11833008, -11690024, -11573026, -11873048, -11373024, -11473002, 
-11703077, -11773029, -11503026, -U1431228, and by the CAS
Key Research Program (KJZDEW-M06) and by Key Research Program of Frontier Sciences, CAS, grant QYZDJ-SSW-SLH007.
JMW thanks the communications with L. Popovi\'c for the polarized spectrum of I Zw 1.
We acknowledge the support of the staff of the Lijiang 2.4m telescope. 
Funding for the telescope has been provided by CAS and the People's Government of Yunnan 
Province.}

\clearpage
\begin{deluxetable}{lcccc}
	\tablecaption{Continuum and H$\beta$ Fluxes}
    \tablecolumns{5}
	\tabletypesize{\footnotesize}
	\tablewidth{0pt}
	\tablehead{
		\colhead{JD-} &
		\colhead{}    &
		\colhead{$F_{5100}$} &
		\colhead{}       &
		\colhead{$F_{\rm H\beta}$}\\
		\colhead{2456700+}   &
		\colhead{}           &
		\colhead{$(10^{-15}\ergscma)$}&
		\colhead{}           &
		\colhead{$(10^{-13}\ergs~{\rm cm^{-2}})$}
	}
\startdata
271.28 &  &6.11 $ \pm $  0.06 & & 2.58 $ \pm $  0.02 \\
274.08 &  &6.44 $ \pm $  0.07 & & 2.75 $ \pm $  0.02 \\
277.27 &  &6.40 $ \pm $  0.08 & & 2.62 $ \pm $  0.02 \\
284.12 &  &6.44 $ \pm $  0.06 & & 2.58 $ \pm $  0.02 \\
286.19 &  &6.44 $ \pm $  0.06 & & 2.64 $ \pm $  0.02 \\	
\enddata
\tablecomments{\footnotesize 
$F_{5100}$ is the flux at $(1+z)$5100 \AA\  in the observed frame.\\
(This table is available in its entirety in machine-readable form.)}
\label{Tab:flux}
\end{deluxetable}

\begin{deluxetable}{lcccccc}
\renewcommand{\arraystretch}{1.5}
    \tablecaption{Characteristics of the Light Curves}
    \tablecolumns{3}
    \tabletypesize{\footnotesize}
    \tablewidth{0pt}
    \tablehead{
        \colhead{Parameter}   &
        \multicolumn{2}{c}{$2014-2015$}  &
        \multicolumn{2}{c}{$2015-2016$}  &
        \multicolumn{2}{c}{$2014-2016$}
    }
\startdata
$N_{5100}$          & \mcn{2}{c}{27}   & \mcn{2}{c}{39} & \mcn{2}{c}{66} \\
$N_{V}$       & \mcn{2}{c}{109}  & \mcn{2}{c}{110}& \mcn{2}{c}{219}\\
$N_{\rm combine}$   & \mcn{2}{c}{136}  & \mcn{2}{c}{149}& \mcn{2}{c}{285}\\
$N_{\rm H_{\beta}}$ & \mcn{2}{c}{27}   & \mcn{2}{c}{39} & \mcn{2}{c}{66} \\
$F_{\rm var}^{5100}(\%)$        & \mcn{2}{c}{$3.3 \pm 0.6 $} & \mcn{2}{c}{$4.1 \pm 0.6 $} & \mcn{2}{c}{$4.3 \pm 0.4 $} \\
$F_{\rm var}^{V}(\%)$        & \mcn{2}{c}{$4.2 \pm 0.4 $} & \mcn{2}{c}{$1.6 \pm 0.3 $} & \mcn{2}{c}{$4.3 \pm 0.3 $} \\
$F_{\rm var}^{\rm combine}(\%)$ & \mcn{2}{c}{$9.3 \pm 0.7 $} & \mcn{2}{c}{$3.7 \pm 0.4 $} & \mcn{2}{c}{$8.5 \pm 0.4 $} \\
$F_{\rm var}^{\rm H \beta}(\%)$ & \mcn{2}{c}{$6.8 \pm 1.0 $} & \mcn{2}{c}{$2.4 \pm 0.4 $} & \mcn{2}{c}{$5.9 \pm 0.5 $} \\                                                                                                                                                                                                                                                                                                                                                                                                                                                                                                                                                                                                                                                                                                                                                                                                                                                                                                                                                                                                                                                                                                                                                                                                                                                                                                                                                                                                                                                                                               \enddata
\end{deluxetable}
\label{Tab:var}

\begin{deluxetable}{lcccccc}
\renewcommand{\arraystretch}{1.5}
    \tablecaption{Results of Correlation Analysis}
    \tablecolumns{3}
    \tabletypesize{\footnotesize}
    \tablewidth{0pt}
    \tablehead{
        \colhead{Parameter}   & 
        \multicolumn{2}{c}{$2014-2015$}  &
        \multicolumn{2}{c}{$2015-2016$}  &
        \multicolumn{2}{c}{$2014-2016$}  \\
        \colhead{}    &
        \colhead{Observed}     &
        \colhead{Rest-frame}   &
        \colhead{Observed}     &
        \colhead{Rest-frame}   &
        \colhead{Observed}     &
        \colhead{Rest-frame}    
    }
\startdata
\mcn{7}{c}{ $\rm H \beta$ vs $V$-band } \\
\hline 
  $r_{\rm max}$                 & 0.91     & 0.91         & 0.73          & 0.73  & 0.88          & 0.88  \\
  $\tau_{\rm cent}$(days)    & 39.5$^{+4.9}_{-6.1}$   & 37.2$^{+4.5}_{-4.9}$   & 48.2$^{+15.2}_{-11.4}$ & 45.4$^{+14.3}_{-10.7}$ & 49.8$^{+4.7}_{-4.0}$ & 46.9$^{+4.5}_{-3.7}$\\
  $\tau_{\rm peak}$(days)    & 27.8$^{+14.1}_{-5.8}$  & 26.2$^{+13.3}_{-5.5}$  & 48.4$^{+19.9}_{-14.8}$ & 45.6$^{+18.8}_{-13.9}$ & 46.2$^{+14.4}_{-7.5}$& 43.6$^{+13.6}_{-7.0}$\\
  $\tau_{\rm JAVELIN}$(days) & 31.6$^{+1.4}_{-0.8}$   & 29.8$^{+1.3}_{-0.8}$   & 53.3$^{+5.4}_{-4.1}$   & 50.2$^{+5.1}_{-3.9}$   & 34.9$^{+7.6}_{-2.5}$ & 32.9$^{+7.2}_{-2.4}$\\
  $\tau_{\rm MICA}$(days)    & 36.7$^{+16.1}_{-16.1}$ & 34.6$^{+15.2}_{-15.2}$ & 50.9$^{+10.3}_{-10.3}$ & 48.0$^{+9.7}_{-9.7}$   & 36.9$^{+2.5}_{-2.5}$ & 34.8$^{+2.4}_{-2.4}$\\  
\enddata
\tablecomments{\footnotesize 
$\tau_{\rm cent}$ and $\tau_{\rm peak}$ are lags from ICCF analysis with the corresponding correlation coefficient 
($r_{\rm max}$), while $\tau_{\rm JAVELIN}$  is the lag from
JAVELIN and $\tau_{\rm MICA}$ is the lag from MICA.  }
\label{Tab:lags}
\end{deluxetable}

\begin{deluxetable}{ll}
\renewcommand{\arraystretch}{1.5}
    \tablecaption{Results of the RM Campaign in the 2014--2015 Season}
    \tablecolumns{2}
    \tabletypesize{\footnotesize}
    \tablewidth{0pt}
    \tablehead{
        \colhead{Parameter}   &
        \colhead{Value}
    }
\startdata   
  $ F_{\rm obs}$(5100 \AA)  & {$6.62 _{-0.28} ^{+0.28} \times 10^{-15} \ \ergscma$} \\
  $ F_{\rm gal}$(5100 \AA)  & {$1.50 _{-0.16} ^{+0.16}\times 10^{-15} \ \ergscma$}\\
  $ F_{\rm AGN}$(5100 \AA)  & {$5.12 _{-0.44} ^{+0.44} \times 10^{-15} \ \ergscma$}\\
  $ L_{5100}$(AGN)         & {$3.19_{-0.27} ^{+0.27} \times 10^{44} \ \ergs$}\\ \hline
  FWHM(mean)           & {$ 1131_{-38}^{+35}\, \kms$}\\
  FWHM(rms)            & {$ 606_{-28}^{+28}\, \kms$}\\
  $ \mbh$                  & {$ 9.30_{-1.38}^{+1.26}\times 10^6\sunm$ (from the mean spectrum)}\\
                           & {$ 2.99_{-0.48}^{+0.46}\times 10^6\sunm$ (from the rms spectrum)}\\
                           & {$ 8.18_{-1.75}^{+1.74}\times 10^6\sunm$ (from the polarized spectrum)}\\
  $\dot{\mathscr{M}}$      & {$203.9^{+61.0}_{-65.8}$} \\
\enddata
\tablecomments{\footnotesize 
$F_{\rm obs}$ and $  F_{\rm gal} $ are the observed and host galaxy flux densities at 
$(1+z)5100$ \AA\ in the observed frame. $L_{5100}$ is the mean AGN luminosity $\lambda L_{\lambda}$ at
rest-frame 5100 $\rm \AA$ after subtracting the host galaxy and correcting for
Galactic extinction. }
\label{Tab:para}
\end{deluxetable}

\clearpage
\appendix
\section{Light curve of the comparison star}
\label{sec:A}
\begin{figure*}
\begin{center}
\includegraphics[angle=0, width=\textwidth]{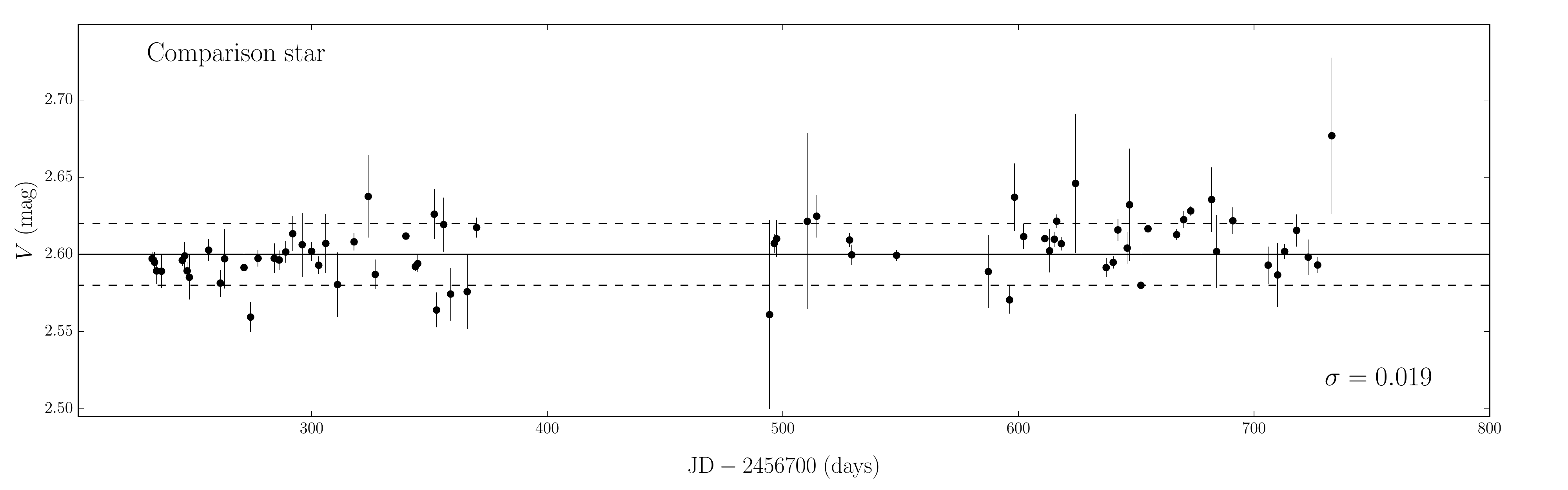}
\end{center}
\vglue -0.5cm
\caption{\footnotesize 
The photometry light curves for the comparison star of I Zw 1. The solid line marks the average 
value and the dashed lines mark the $\pm 1\sigma$ standard deviation.}
\label{compare}
\end{figure*}

In order to check the stability of the comparison star during the 
campaign, we obtain its photometric light curve using five stars in the same field to perform 
differential photometry. The light curve of the comparison star is shown in Figure \ref{compare}. 
The standard deviation is 1.9$\%$ , demonstrating that the comparison star does not vary significantly.

\section{Light curves and H$\beta$ lags}
\label{sec:B}
\begin{figure*}[t!]
\begin{center}
\includegraphics[angle=0, width=\textwidth]{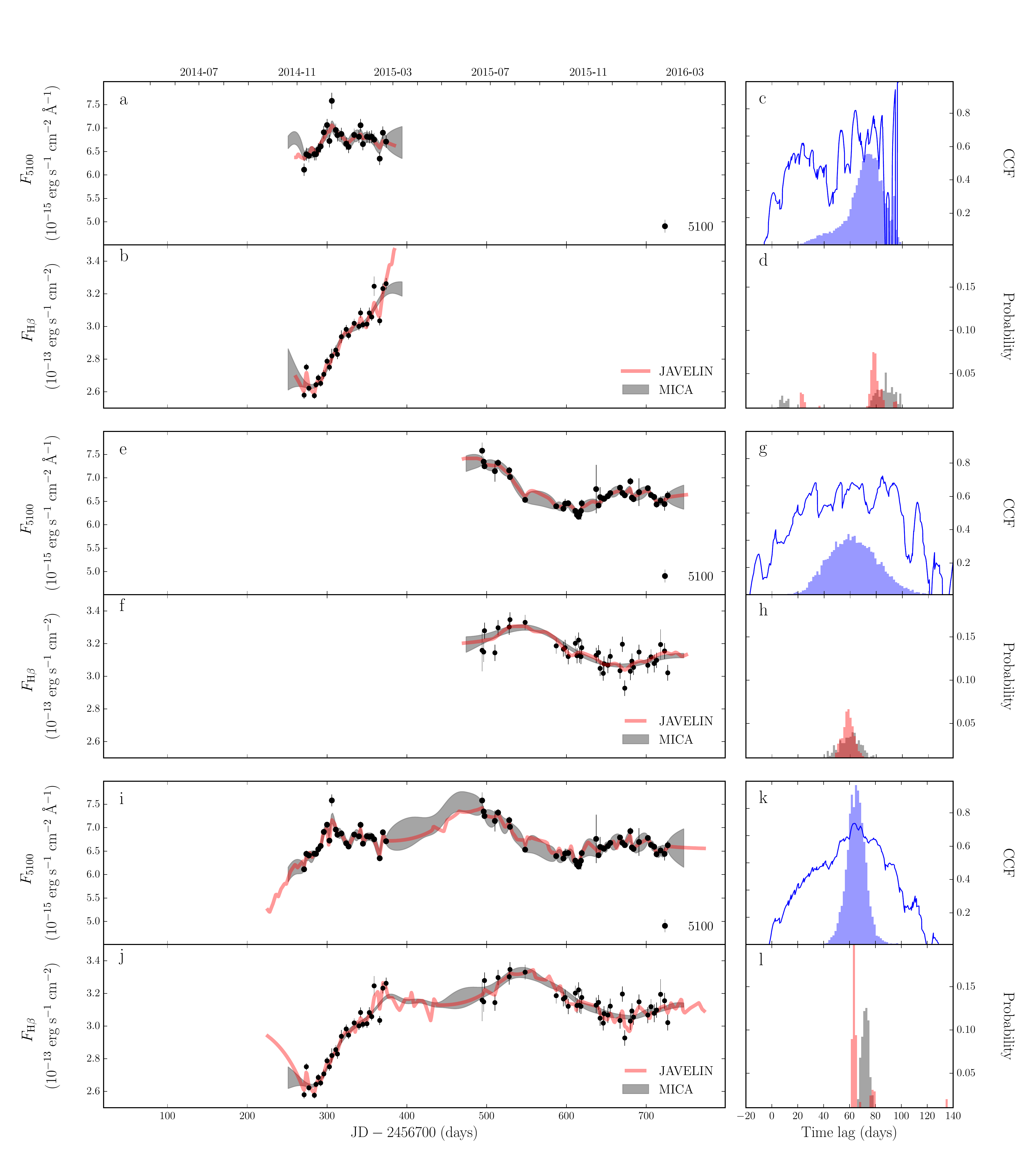}
\end{center}
\caption{\footnotesize 
Light curves and results of correlation analysis using the 5100 \AA\ continuum light curves for I Zw 1 (same as Figure \ref{Fig:photo}).
}
\label{Fig:f5100}
\end{figure*}

\begin{figure*}[t!]
\begin{center}
\includegraphics[angle=0, width=\textwidth]{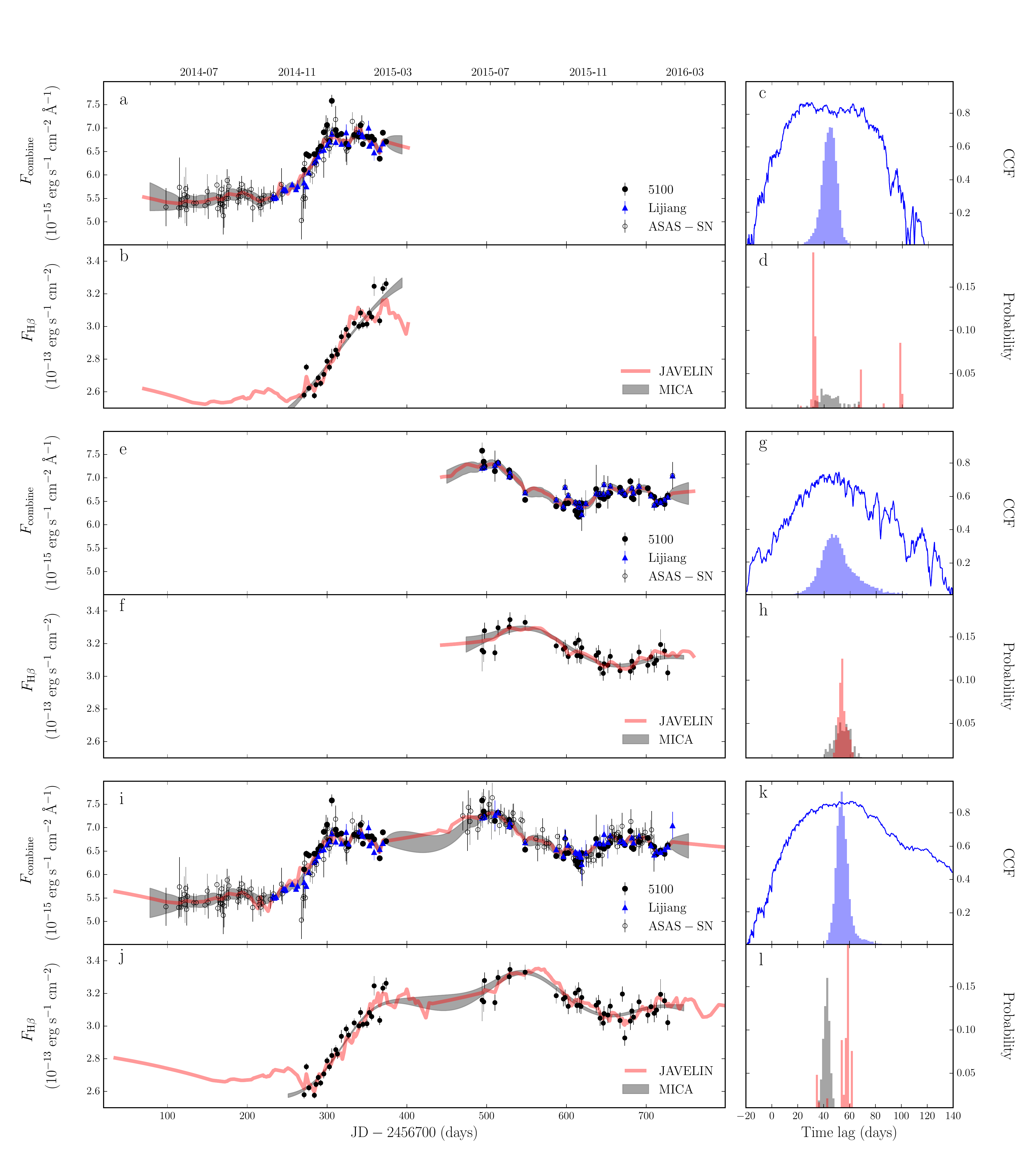}
\end{center}
\caption{\footnotesize 
Light curves and results of correlation analysis using the combined continuum light curves for I Zw 1 (same as Figure \ref{Fig:photo}).
}
\label{Fig:fcombine}
\end{figure*}

Figure \ref{Fig:f5100} and \ref{Fig:fcombine} show the results of the correlation analysis 
using the 5100 \AA\ and combined continuum light curves. 
For the 2014--2015 data, the H$\beta$ time lag cannot be constrained well solely using the 5100 \AA\ 
light curves. As shown in Figures \ref{Fig:f5100}c and 
\ref{Fig:f5100}d, the CCF is very spiky and the posterior distributions of the time lags obtained by 
JAVELIN and MICA are not well-defined. This may be caused by the short duration and the large scatter 
of the continuum light curves. As shown in Tables \ref{Tab:B1} and \ref{Tab:lags}, the uncertainty of the time lags 
derived using only the photometry continuum light curves is smaller than that using the combined continuum 
light curves. 

\begin{figure*}[t!]
\begin{center}
\includegraphics[angle=0, width=\textwidth]{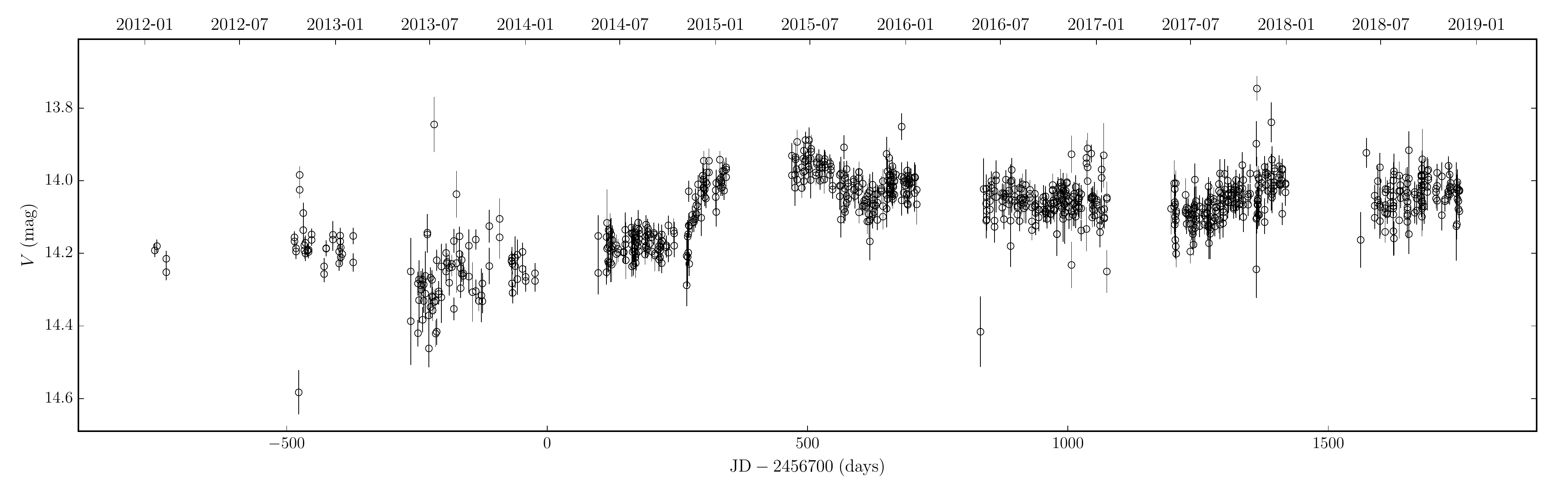}
\end{center}
\caption{\footnotesize 
The 7-year $V-$band light curves of I Zw 1 monitored by ASAS-SN. It shows that I Zw 1 is quite stable in
long-term. This makes it difficult to capture reverberation of the H$\beta$ line.
}
\label{Fig:7year}
\end{figure*}

In Figure \ref{Fig:7year}, we show the ASAS-SN light curve of I Zw 1
from 2012 to the end of 2018. 
Generally, the source was less variable in those 7 years except for the period from 2014-07 to 2016-01.  During the whole period, the  variability of the continuum flux is $F_{\rm var}(\%)=0.63\pm0.02$,
which is much smaller than that of the 2014-2016 period. We were lucky in this SEAMBH campaign and captured the largest variations in this period. Figure \ref{Fig:7year}
shows that it is less variable again after the present campaign, which implys that we may need a longer waiting for sharp variations.
 Higher cadence than 2-3 days is also necessary for more details of the BLR structure. 
On the other hand, the long-term stability of I Zw 1 supports SEAMBHs as cosmic candles for cosmology \citep{wang2013,Wang2014,Cai2018}. 

\begin{deluxetable}{lcccccc}
\renewcommand{\arraystretch}{1.5}
    \tablecaption{Results of Correlation Analysis}
    \tablecolumns{3}
    \tabletypesize{\footnotesize}
    \tablewidth{0pt}
    \tablehead{
        \colhead{Parameter}   & 
        \multicolumn{2}{c}{$2014-2015$}  &
        \multicolumn{2}{c}{$2015-2016$}  &
        \multicolumn{2}{c}{$2014-2016$}  \\
        \colhead{}    &
        \colhead{Observed}     &
        \colhead{Rest-frame}   &
        \colhead{Observed}     &
        \colhead{Rest-frame}   &
        \colhead{Observed}     &
        \colhead{Rest-frame}    
    }
\startdata
\mcn{7}{c}{$\rm H \beta$ vs $F_{5100}$} \\
\hline
  $r_{\rm max}$                 & --    &--         &0.72           & 0.72 & 0.74 & 0.74 \\
  $\tau_{\rm cent}$(days)    & 73.3$^{+11.7}_{-14.8}$ & 69.1$^{+11.0}_{-13.9}$ & 61.9$^{+19.8}_{-18.2}$ & 58.3$^{+18.7}_{-17.2}$ & 65.2$^{+6.9}_{-7.0}$  & 61.4$^{+6.5}_{-6.6}$\\
  $\tau_{\rm peak}$(days)    & 87.7$^{+8.9}_{-20.4}$  & 82.7$^{+8.4}_{-19.2}$  & 57.9$^{+33.5}_{-24.0}$ & 54.6$^{+31.6}_{-22.6}$ & 64.0$^{+7.9}_{-4.1}$  & 60.3$^{+7.4}_{-3.8}$\\
 $\tau_{\rm JAVELIN}$(days)  & 78.0$^{+7.4}_{-47.8}$  & 73.5$^{+7.0}_{-45.0}$  & 59.1$^{+10.4}_{-7.0}$  & 55.7$^{+9.8}_{-6.6}$ & 64.0$^{+13.3}_{-1.2}$ & 60.3$^{+12.5}_{-1.1}$\\
  $\tau_{\rm MICA}$(days)    & 66.2 $^{+34.9}_{-34.9}$& 62.4$^{+32.9}_{-32.9}$ & 59.9$^{+12.5}_{-12.5 }$& 56.5$^{+11.8}_{-11.8}$ & 72.1$^{+2.9}_{-2.9}$& 67.9$^{+2.7}_{-2.7}$\\ 
\hline
\hline 
\mcn{7}{c}{ $\rm H \beta$ vs $F_{\rm combine}$ } \\
\hline 
  $r_{\rm max}$                 & 0.86     & 0.86         & 0.75          & 0.75 & 0.87 & 0.87 \\
  $\tau_{\rm cent}$(days)    & 44.1$^{+5.1}_{-5.5}$   & 41.6$^{+4.8}_{-5.2}$   & 48.9$^{+15.0}_{-10.4}$ & 46.1$^{+14.1}_{-9.8}$ & 54.2$^{+5.0}_{-4.1}$ & 51.1$^{+4.6}_{-3.9}$\\
  $\tau_{\rm peak}$(days)    & 39.3$^{+22.7}_{-13.0}$ & 37.6$^{+21.4}_{-12.3}$ & 49.2$^{+18.7}_{-14.4}$ & 46.4$^{+17.6}_{-13.6}$& 57.4$^{+7.0}_{-13.1}$ & 54.1$^{+6.6}_{-12.4}$\\
  $\tau_{\rm JAVELIN}$(days) & 33.1$^{+65.2}_{-1.5}$  & 31.2$^{+61.4}_{-1.4}$  & 54.0$^{+5.4}_{-3.0}$   & 50.9$^{+4.8}_{-2.8}$ & 57.9$^{+1.0}_{-6.5}$ & 54.6$^{+1.0}_{-6.1}$\\
  $\tau_{\rm MICA}$(days)    & 45.5$^{+22.8}_{-22.8}$ & 42.9$^{+21.5}_{-21.5}$ & 53.6$^{+10.5}_{-10.5 }$& 50.5$^{+9.9}_{-9.9}$ & 42.2$^{+2.7}_{-2.7}$  & 39.8$^{+2.5}_{-2.5}$ \\
\enddata
\tablecomments{\footnotesize 
$\tau_{\rm cent}$ and $\tau_{\rm peak}$ are lags from ICCF analysis with the corresponding correlation coefficient 
($r_{\rm max}$), while $\tau_{\rm JAVELIN}$ is the lag for
JAVELIN and $\tau_{\rm MICA}$ is the lag for MICA. }
\label{Tab:B1}
\end{deluxetable}


\begin{thebibliography}{99}

\bibitem[Abramowicz et al. (1988)]{Abramowicz1988}Abramowicz, M., Czerny, B., Lasota, J. \& Szuszkiewicz, E. 1988, \apj, 332, 646
\bibitem[Ade et al.(2014)]{ade14}Ade, P.A.R., Arnaud, M., et al.(Planck Collaboration)2014, \aap , 571.A31
\bibitem[Afanasiev \& Popovic (2015)]{Afanasiev2015}Afanasiev, V. L. \& Popovi\'c, L. 2015, \apj, 800, L35
\bibitem[Baldassare et al.(2017)]{Baldassare2017} Baldassare, V.~F., Reines, A.~E., Gallo, E., \& Greene, J.~E.\ 2017, \apj, 850, 196
\bibitem[Baldi et al.(2016)]{baldi16}Baldi, R.~D., Capetti, A., Robinson, A., Laor, A., \& Behar, E.\ 2016, \mnras, 458, L69
\bibitem[Bamnados et al. (2018)]{Banados2018}Banados, E. et al. 2018, \nat, 553, 473
\bibitem[Bell \& de Jong(2001)]{Bell2001}Bell, E.~F., \& de Jong, R.~S.\ 2001, \apj, 550, 212
\bibitem[Bentz et al.(2013)]{bentz13}Bentz, M.~C., Denney, K.~D., Grier, C.~J., et al.\ 2013, \apj, 767, 149
\bibitem[Binney \& Merrifield(1998)]{Binney1998}Binney, J., \& Merrifield, M.\ 1998, Galactic astronomy / James Binney and Michael Merrifield.~ Princeton, NJ : Princeton University Press, 1998.~ (Princeton series in astrophysics) QB857 .B522 1998
\bibitem[Blandford \& McKee(1982)]{bland82}Blandford, R.~D., \& McKee, C.~F.\ 1982, \apj, 255, 419
\bibitem[Boller et al.(1996)]{boller96}Boller, T., Brandt, W.~N., \& Fink, H.\ 1996, \aap, 305, 53
\bibitem[Boroson \& Green(1992)]{bor92}Boroson, T.~A., \& Green, R.~F.\ 1992, \apjs, 80, 109
\bibitem[Bruzual \& Charlot(2003)]{Bruzual2003}Bruzual, G., \& Charlot, S.\ 2003, \mnras, 344, 1000
\bibitem[Cai et al. (2018)]{Cai2018}Cai, R.-G., Guo, Z.-K., Huang, Q.-G. \& Yang, T. 2018, \prd, 97, 123502
\bibitem[Cardelli et al.(1989)]{cardelli1989}Cardelli, J.~A., Clayton, G.~C., \& Mathis, J.~S.\ 1989, \apj, 345, 245
\bibitem[Chabrier(2003)]{Chabrier2003}Chabrier, G.\ 2003, \pasp, 115, 763
\bibitem[Collin et al.(2006)]{collin06}Collin, S., Kawaguchi, T., Peterson, B.~M., \& Vestergaard, M.\ 2006, \aap, 456, 75
\bibitem[Du et al.(2014)]{du14}Du, P., Hu, C., Lu, K.-X., et al.\ 2014, \apj, 782, 45
\bibitem[Du et al.(2015)]{du15}Du, P., Hu, C., Lu, K.-X., et al.\ 2015, \apj, 806, 22
\bibitem[Du et al.(2016a)]{du16a} Du, P., Lu, K.-X., Zhang, Z.-X., et al.\ 2016a, \apj, 825, 126 
\bibitem[Du et al.(2016b)]{du16b}Du, P., Wang, J.-M., Hu, C., et al.\ 2016b, \apjl, 818, L14
\bibitem[Du et al.(2017)]{du17}Du, P., Wang, J.-M., \& Zhang, Z.-X.\ 2017, \apjl, 840, L6
\bibitem[Du et al.(2018)]{du18}Du, P., Zhang, Z.-X., Wang, K., et al.\ 2018, \apj, 856, 6
\bibitem[Edelson et al. (2002)]{Edelson2002}Edelson, R., Turner, T. J., Pounds, K. et al. 2002, \apj, 568, 610
\bibitem[Fisher \& Drory(2008)]{Fisher2008}Fisher, D.~B., \& Drory, N.\ 2008, \aj, 136, 773 
\bibitem[Foltz et al.(1981)]{foltz81}Foltz, C.~B., Peterson, B.~M., Cariotti, E.~R., et al.\ 1981, \apj, 250, 508 
\bibitem[Gadotti (2009)]{Gadotti2009}Gadotti, D. A. 2009, \mnras,
\bibitem[Gallo et al.(2007)]{gallo07}Gallo, L.~C., Brandt, W.~N., Costantini, E., \& Fabian, A.~C.\ 2007, \mnras, 377, 1375
\bibitem[Gao \& Ho(2017)]{gao2017}Gao, H., \& Ho, L.~C.\ 2017, \apj, 845, 114 
\bibitem[Gaskell \& Sparke(1986)]{gas86}Gaskell, C.~M., \& Sparke, L.~S.\ 1986, \apj, 305, 175
\bibitem[Giannuzzo et al.(1998)]{gian98}Giannuzzo, M.~E., Mignoli, M., Stirpe, G.~M., \& Comastri, A.\ 1998, \aap, 330, 894
\bibitem[Goodrich(1989)]{good89}Goodrich, R.~W.\ 1989, \apj, 342, 224
\bibitem[Graham et al.(2011)]{graham11}Graham, A.~W., Onken, C.~A., Athanassoula, E., \& Combes, F.\ 2011, \mnras, 412, 2211
\bibitem[Greene \& Ho (2005)]{Greene2005}Grenne, J. \& Ho, L. C. 2005, \apj, 630, 122
\bibitem[Grupe(2004)]{grupe04}Grupe, D.\ 2004, \aj, 127, 1799
\bibitem[Halpern \& Oke(1987)]{hal87}Halpern, J.~P., \& Oke, J.~B.\ 1987, \apj, 312, 91
\bibitem[Ho \& Kim(2014)]{ho14}Ho, L.~C., \& Kim, M.\ 2014, \apj, 789, 17
\bibitem[Hu(2009)]{hu09}Hu, J.\ 2009, arXiv:0908.2028 
\bibitem[Hu et al. (2008)]{hu08}Hu, C., Wang, J.-M., Ho, L.~C., et al.\ 2008, \apj, 687, 78-96 
\bibitem[Hu et al.(2012)]{hu12}Hu, C., Wang, J.-M., Ho, L.~C., et al.\ 2012, \apj, 760, 126
\bibitem[Hu et al. (2015)]{hu15}Hu, C., Du, P., Lu, K.-X. et al. 2015, \apj, 804, 138
\bibitem[Hu et al.(2016)]{hu16} Hu, C., Wang, J.-M., Ho, L.~C., et al.\ 2016, \apj, 832, 197 
\bibitem[Hutchings \& Crampton(1990)]{hut90}Hutchings, J.~B., \& Crampton, D.\ 1990, \aj, 99, 37
\bibitem[Kaspi et al.(2000)]{kas00}Kaspi, S., Smith, P.~S., Netzer, H., et al., \ 2000, \apj, 533, 631
\bibitem[Kawaguchi et al.(2004)]{Kawaguchi2004}Kawaguchi, T., Aoki, K., Ohta, K. \& Collin, S. 2004, \aap, 420, L23
\bibitem[Kelly et al.(2009)]{kel09}Kelly, B.~C., Bechtold, J., \& Siemiginowska, A., \ 2009, \apj, 698, 895-910
\bibitem[Kim et al.(2008)]{Kim2008a} Kim, M., Ho, L.~C., Peng, C.~Y., Barth, A.~J., \& Im, M.\ 2008, \apjs, 179, 283 
\bibitem[Kim et al.(2008)]{Kim2008b}Kim, M., Ho, L.~C., Peng, C.~Y., et al.\ 2008, \apj, 687, 767 
\bibitem[Kim et al.(2017)]{Kim2017}Kim, M., Ho, L.~C., Peng, C.~Y., Barth, A.~J., \& Im, M.\ 2017, \apjs, 232, 21 
\bibitem[Kinney et al.(1996)]{kinn96}Kinney, A.~L., Calzetti, D., Bohlin, R.~C., et al.\ 1996, \apj, 467, 38
\bibitem[Kochanek et al.(2017)]{Kochanek2017}
Kochanek, C. S., Shappee, B. J., Stanek, K. Z., et al. 2017, \pasp, 129, 104502
\bibitem[Kormendy \& Ho(2013)]{Kormendy2013}Kormendy, J., \& Ho, L.~C.\ 2013, \araa, 51, 511
\bibitem[Kormendy \& Richstone(1995)]{kor95}Kormendy, J., \& Richstone, D.\ 1995, \araa, 33, 581
\bibitem[Krist \& Hook(1999)]{Krist1999}
Krist, J.,\& Hook, R.1999, The Tiny Time User's  Guide (Baltimore: STScI)
\bibitem[Krolik (2001)]{Krolik2001}Krolik, J. H. 2001, \apj, 551, 72
\bibitem[Li et al.(2013)]{li13}Li, Y.-R., Wang, J.-M., Ho, L.~C., Du, P., \& Bai, J.-M.\ 2013, \apj, 779, 110 
\bibitem[Li et al.(2014)]{li14} Li, Y.-R., Wang, J.-M., Hu, C., Du, P., \& Bai, J.-M.\ 2014, \apjl, 786, L6 
\bibitem[Li et al.(2016)]{li16}Li, Y.-R., Wang, J.-M., \& Bai, J.-M.\ 2016, \apj, 831, 206
\bibitem[Li et al.(2018)]{Li2018}
Li, Y.-R., Songsheng, Y.-Y., Qiu, J. et al. (SEAMBH collaboration), 2018, \apj, 869, 137
\bibitem[Longhetti \& Saracco(2009)]{Longhetti2009}Longhetti, M., \& Saracco, P.\ 2009, \mnras, 394, 774
\bibitem[Maoz et al.(1990)]{maoz90}Maoz, D., Netzer, H., Leibowitz, E., et al.\ 1990, \apj, 351, 75
\bibitem[Marziani \& Sulentic (2014)]{Marziani2014}Marziani, P. \& Sulentic, J. 2014, \mnras, 442, 1211
\bibitem[Marziani et al.(2019)]{Marziani2019}
 Marziani, P., Bon, E., Bon, N. et al. arXiv:1901.10032
\bibitem[Mathur (2000)]{Mathur2000}Mathur, S. 2000, \mnras, 314, L17
\bibitem[Mathur \& Grupe(2005)]{mathur05}Mathur, S., \& Grupe, D.\ 2005, \apj, 633, 688
\bibitem[Mart\'inez-Aldama et al(2018)]{Mary2018}
Mart\'inez-Aldama, M. L.; del Olmo, A.; Marziani, P. et al. 2018, \aap, 618, 179
\bibitem[McConnell \& Ma(2013)]{mcc13}McConnell, N.~J., \& Ma, C.-P.\ 2013, \apj, 764, 184
\bibitem[Mineshige et al. (2000)]{Mineshige2000}Mineshige, S., Kawaguchi, T., Takeuchi, M. \& Hayashida, K 2000, \pasj, 52, 499
\bibitem[Mortlock et al. (2011)]{Mortlock2011}Mortlock, D. J. et al. 2011, \nat, 474, 616
\bibitem[Negrete et al. (2018)]{Negret2018}
 Negret,  C. A., Dultzin, D.,  Marziani, P. Et al. 2018, \aap, 620, 118
\bibitem[Netzer \& Marziani(2010)]{netzer10}Netzer, H., \& Marziani, P.\ 2010, \apj, 724, 318 
\bibitem[Negrete et al.(2012)]{neg12}Negrete, C.~A., Dultzin, D., Marziani, P., \& Sulentic, J.~W.\ 2012, \apj, 757, 62
\bibitem[Onken et al.(2004)]{onken04}Onken, C.~A., Ferrarese, L., Merritt, D., et al.\ 2004, \apj, 615, 645
\bibitem[Osterbrock \& Mathews(1986)]{ost86}Osterbrock, D.~E., \& Mathews, W.~G.\ 1986, \araa, 24, 171
\bibitem[Osterbrock \& Pogge(1985)]{ost85}Osterbrock, D.~E., \& Pogge, R.~W.\ 1985, \apj, 297, 166
\bibitem[Pancoast et al.(2014)]{pan14}Pancoast, A., Brewer, B.~J., Treu, T., et al.\ 2014, \mnras, 445, 3073
\bibitem[Peng et al.(2002)]{Peng2002}Peng, C.~Y., Ho, L.~C., Impey, C.~D., \& Rix, H.-W.\ 2002, \aj, 124, 266
\bibitem[Peng et al.(2010)]{Peng2010}Peng, C.~Y., Ho, L.~C., Impey, C.~D., \& Rix, H.-W.\ 2010, \aj, 139, 2097 
\bibitem[Peterson et al.(1982)]{peterson1982}Peterson, B.~M., Foltz, C.~B., Byard, P.~L., \& Wagner, R.~M.\ 1982, \apjs, 49, 469 
\bibitem[Peterson (1993)]{Peterson1993}Peterson, B. 1993, \pasp, 105, 247
\bibitem[Peterson et al.(1998)]{peter98}Peterson, B.~M., Wanders, I., Horne, K., et al.\ 1998, \pasp, 110, 660
\bibitem[Peterson et al.(2004)]{peter04}Peterson, B.~M., Ferrarese, L., Gilbert, K.~M., et al.\ 2004, \apj, 613, 682
\bibitem[Peterson (2014)]{Peterson2014}Peterson, B. 2014, SSRv, 183, 253
\bibitem[Rodr{\'{\i}}guez-Pascual et al.(1997)]{rod97}Rodr{\'{\i}}guez-Pascual, P.~M., Alloin, D., Clavel, J., et al.\ 1997, \apjs, 110, 9
\bibitem[Salpeter(1955)]{Salpeter1955}Salpeter, E.~E.\ 1955, \apj, 121, 161
\bibitem[Sargent(1968)]{sar68}Sargent, W.~L.~W.\ 1968, \apjl, 152, L31
\bibitem[Scharw{\"a}chter et al.(2003)]{schar03}Scharw{\"a}chter, J., Eckart, A., Pfalzner, S., et al.\ 2003, \aap, 405, 959
\bibitem[Scharw{\"a}chter et al.(2007)]{Scharwachter2007}Scharw{\"a}chter, J., Eckart, A., Pfalzner, S., Saviane, I., \& Zuther, J.\ 2007, \aap, 469, 913
\bibitem[Schlafly \& Finkbeiner(2011)]{schlafly11}Schlafly, E.~F., \& Finkbeiner, D.~P.\ 2011, \apj, 737, 103
\bibitem[Schmidt \& Green(1983)]{schm83}Schmidt, M., \& Green, R.~F.\ 1983, \apj, 269, 352
\bibitem[\sersic(1968)]{sersic1968}\sersic, J.~L.\ 1968, Cordoba, Argentina: Observatorio Astronomico
\bibitem[Shakura \& Sunyeav (1973)]{Shakura1973}Shakura, N. I. \& Sunyaev, R. 1973, \aap, 24, 337
\bibitem[Shappee et al.(2014)]{Shappee2014}
Shappee, B. J., Prieto, J. L., Grupe, D., et al. 2014, \apj, 788, 48
\bibitem[Slavcheva-Mihova \& Mihov(2011a)]{sla11a}Slavcheva-Mihova, L., \& Mihov, B.\ 2011a, \aap, 526, A43
\bibitem[Slavcheva-Mihova \& Mihov(2011b)]{sla11}Slavcheva-Mihova, L., \& Mihov, B.\ 2011b, Astronomische Nachrichten, 332, 191
\bibitem[Smith et al.(1997)]{smith97}Smith, P.~S., Schmidt, G.~D., Allen, R.~G., \& Hines, D.~C.\ 1997, \apj, 488, 202
\bibitem[Smith et al.(2002)]{smith02}Smith, J.~E., Young, S., Robinson, A., et al.\ 2002, \mnras, 335, 773
\bibitem[Smith et al. (2005)]{Smith2005}Smith, J. E., Robinson, A., Young, S., Axon, D. J. \& Corbett, E.
A. 2005, \mnras, 359, 846
\bibitem[Songsheng \& Wang (2018)]{Songsheng2018}Songsheng, Y.-Y. \& Wang, J.-M. 2018, \mnras, 473, L1
\bibitem[Sulentic et al.(2000)]{sulen00}Sulentic, J.~W., Marziani, P., Zwitter, T., Dultzin-Hacyan, D., \& Calvani, M.\ 2000, \apjl, 545, L15
\bibitem[Sturm et al.(2018)]{Sturm2018}
Strum, E., Dexter, J., Pfuhl, O. et al. 2018, \nat, 563, 657
\bibitem[van Dokkum(2001)]{vand01}van Dokkum, P.~G.\ 2001, \pasp, 113, 1420
\bibitem[Vestergaard \& Peterson(2006)]{vester06}Vestergaard, M., \& Peterson, B.~M.\ 2006, \apj, 641, 689 \bibitem[V{\'e}ron-Cetty et al.(2004)]{vero04}V{\'e}ron-Cetty, M.-P., Joly, M., \& V{\'e}ron, P.\ 2004, \aap, 417, 515
\bibitem[Vestergaard \& Wilkes(2001)]{vest01}Vestergaard, M., \& Wilkes, B.~J.\ 2001, \apjs, 134, 1
\bibitem[Volonteri \& Rees (2005)]{Volonteri2005}Volonteri, M. \& Rees, M. J. 2005, \apj, 633, 624
\bibitem[Wang \& Zhou(1999)]{wang99}Wang, J.-M., \& Zhou, Y.-Y.\ 1999, \apj, 516, 420
\bibitem[Wang \& Netzer(2003)]{wang2003}Wang, J.-M., \& Netzer, H.\ 2003, \aap, 398, 927
\bibitem[Wang et al. (2004)]{Wang2004}Wang, J.-M., Watarai, K. \&  Mineshige, S. 2004, \apj, 607, L107
\bibitem[Wang et al. (2006)]{Wang2006}Wang, J.-M., Chen, Y.-M. \& Zhang, F. 2006, \apj, 647, L17
\bibitem[Wang \& Zhang (2007)]{Wang2007}Wang, J.-M. \& Zhang, E.-P. 2007, \apj, 660, 1072
\bibitem[Wang et al.(2013)]{wang2013}Wang, J.-M., Du, P., Valls-Gabaud, D., Hu, C., \& Netzer, H.\ 2013, \prl, 110, 081301
\bibitem[Wang et al. (2014)]{Wang2014}Wang, J.-M., Du, P., Hu, C. et al. 2014, \apj, 793, 108
\bibitem[Wang et al. (2014b)]{Wang2014b}Wang, J.-M., Qiu, J., Du, P. \& Ho, L. C. 2014, \apj, 797, 65
\bibitem[Woo et al.(2013)]{woo13}Woo, J.-H., Schulze, A., Park, D., et al.\ 2013, \apj, 772, 49 
\bibitem[Woo et al.(2015)]{woo15} Woo, J.-H., Yoon, Y., Park, S., Park, D., \& Kim, S.~C.\ 2015, \apj, 801, 38 
\bibitem[Zhang \& Wang (2006)]{Zhang2006}Zhang, E.-P. \& Wang, J.-M. 2006, \apj, 653, 137
\bibitem[Zhang et al.(2018)]{Zhang2018}
Zhang, Z.-X.,Du, P., Smith, P. et al. 2018, ApJ, arXiv:1811.03812
\bibitem[Zu et al.(2011)]{zu11}Zu, Y., Kochanek, C.~S., \& Peterson, B.~M., \ 2011, \apj, 735, 80
\end{thebibliography}
\end{document}